\documentclass[12pt,preprint]{aastex}

\newcommand{\myemail}{eduardo.gonzalez@uah.es}
\newcommand{\kms}{{\hbox {km\thinspace s$^{-1}$}}}
\newcommand{\Lsun}{{\hbox {L$_\odot$}}}
\newcommand{\Msun}{{\hbox {M$_\odot$}}}
\newcommand{\Mdot}{{\hbox {$\dot{M}$}}}
\newcommand{\arcmdot}{\rlap{.}'}
\newcommand{\arcsdot}{\rlap{.}''}

\newcommand{\hdo}{{\hbox {H$_{2}$O}}}

\def\t#1#2#3#4#5#6{{\hbox {$#1_{#2#3}\!\rightarrow\!#4_{#5#6}$}}}
\def\13co{$^{13}$CO}
\def\c18o{C$^{18}$O}

\shorttitle{Water vapor emission from IRC+10216 and other carbon-rich stars}
\shortauthors{Gonz\'alez-Alfonso, Neufeld, \& Melnick}

\begin{document}


\title{Water vapor emission from IRC+10216 and other carbon-rich stars:
model predictions and prospects for multitransition observations}



\author{Eduardo Gonz\'alez-Alfonso}
\affil{Universidad de Alcal\'a de Henares, Departamento de F\'{\i}sica, 
Campus Universitario, E-28871 Alcal\'a de Henares, Madrid, Spain}
\email{\myemail}

\author{David A. Neufeld}
\affil{Department of Physics \& Astronomy, The Johns Hopkins University, 
3400 N. Charles Street, Baltimore, Maryland 21218}

\and

\author{Gary J. Melnick}
\affil{Harvard-Smithsonian Center for Astrophysics,
    60 Garden Street, Cambridge, MA 02138}




\begin{abstract}

We have modeled the emission of \hdo\ rotational lines from the
extreme C-rich star IRC+10216. Our treatment of the excitation of
\hdo\ emissions takes into account the excitation of \hdo\ both through 
collisions, and through the pumping of the $\nu_2$ and $\nu_3$ vibrational 
states by dust emission and subsequent decay to the ground state. 
Regardless of the spatial distribution of the water molecules,
the \hdo\ $1_{10}-1_{01}$ line at 557 GHz observed by the 
{\em Submillimeter Wave Astronomy Satellite} (SWAS) is found to be 
pumped primarily through the absorption of dust-emitted 
photons at 6 $\mu$m in the $\nu_2$ band. As noted by previous authors,
the inclusion of radiative pumping lowers the ortho-\hdo\ abundance 
required to account for the 557 GHz emission, which is found
to be $(0.5-1)\times10^{-7}$ if the presence of \hdo\ 
is a consequence of vaporization of orbiting comets or Fischer-Tropsch 
catalysis. Predictions for other submillimeter \hdo\ lines that can be 
observed by the {\em Herschel Space Observatory} (HSO) are reported. 
 Multitransition HSO observations promise to reveal the spatial
distribution of the circumstellar water vapor, discriminating among the 
several hypotheses that have been proposed for the origin of the \hdo\ 
vapor in the envelope of IRC+10216. We also show that, 
for observations with HSO, the \hdo\ $1_{10}-1_{01}$ 557 GHz line 
affords the greatest sensitivity in searching for \hdo\ in other 
C-rich AGB stars.

\end{abstract}




\section{Introduction}
\label{sec:intro}

The discovery of water vapor emission in the extreme C-rich AGB star
IRC+10216 with SWAS \citep{mel01}, its confirmation by ODIN 
\citep{has06}, and the subsequent detection of other
O-bearing molecules like OH \citep{for03}, H$_2$CO \citep{for04},
and C$_3$O \citep{ten06},
have challenged our current understanding of the chemistry in envelopes around
C-rich AGB stars. According to standard models, essentially all oxygen
nuclei are predicted to be locked into CO or SiO with no reservoir 
to form other O-bearing molecules (except for low abundances of
species such as HCO$^+$ in the outer envelope where photochemistry is
important). Therefore, the unexpectedly high abundances found for 
H$_2$O, OH, and H$_2$CO, indicate that several processes not included in 
standard models for C-rich environments are driving the oxygen chemistry,
but the dominant water production mechanism is still a source of debate.

Four distinct mechanisms have been considered as possible sources
of the observed water vapor in IRC+10216, each one making a specific 
prediction for the \hdo\ spatial distribution in the envelope:
$(i)$ chemistry in the inner envelope, which would imply the presence of 
\hdo\ in the warmest and densest regions;  $(ii)$ vaporization of 
icy orbiting bodies that have survived from the main sequence into the C-rich 
AGB phase \citep{mel01,for01}, predicting the release of \hdo\ at 
intermediate radii of $R_{int}=(1-5)\times10^{15}$ cm; $(iii)$ grain surface 
reactions, such as the Fischer-Tropsch catalysis on the surfaces 
of small metallic 
grains \citep{wil04}, which predicts \hdo\ to attain a nearly uniform 
abundance at radii larger than $R_{int}=(1.5-2)\times10^{15}$ cm; $(iv)$ 
chemistry involving photodissociation products in the outermost 
regions of the envelope: a specific mechanism relying on the radiative 
association O+H$_2$$\rightarrow$\hdo+$\gamma$ has been proposed by 
Ag\'undez \& Cernicharo (2006, hereafter AC06),
and predicts \hdo\ to be present at 
radii higher than $R_{int}\approx4\times10^{16}$ cm. In all cases, \hdo\ 
is expected to have
a uniform abundance from $R_{int}$ up to the external region where it is
photodissociated producing OH.

The only water line detected so far in IRC+10216, the ortho-\hdo\ 
\t110101\ transition at 557 GHz, is the ground-state transition, with the 
upper level at only 27 K above the ground rotational level, and cannot 
discriminate -by itself- between the processes listed above. 
Nevertheless, the launch of the Herschel Space Observatory (HSO) will allow 
us to observe other \hdo\ lines in the submillimeter and far-infrared 
domains, permitting us to infer the region where the \hdo\ emission 
arises, and will potentially favor one of the proposed formation 
mechanisms. The Heterodyne Instrument for the Far Infrared (HIFI) onboard 
HSO will provide very high spectral resolution observations (0.14-1.0 MHz), 
thus permitting the lines to be velocity resolved. 
In this paper, we model the \hdo\ emission 
from IRC+10216 to show how the \hdo\ spatial distribution can be 
inferred from HSO multi-transition observations. Also, we explore
which \hdo\ transition provides the most sensitive means of searching for
water vapor around AGB stars other than IRC+10216; such a search would
determine whether the occurence of \hdo\ in C-rich environments is
widespread.

\section{Model description} \label{sec:model}

Our model for IRC+10216 assumes a stellar mass
loss rate of \Mdot$=3\times10^{-5}$ \Msun\ yr$^{-1}$. The gas velocity
field, turbulent velocity, and H$_2$ density and gas temperature profiles 
are those adopted by \cite{mel01} (see references therein), and listed 
in Table~\ref{tab:envel}. A distance to the source $D=170$ pc is assumed. 

The adopted gas temperature 
profile is an approximate fit \citep[e.g.][]{gla96} to the results obtained 
by \cite{kwa82}, who calculated the thermal profile through the circumstellar 
envelope of IRC+10216 from the energetic balance between gas heating and 
cooling. These calculations included gas-dust collisional heating and both
adiabatic and CO line cooling. HCN line cooling was ignored and, subsequently,
the ISO far-infrared spectrum of IRC+10216 showed strong HCN rotational
emission from both the ground and excited vibrational states \citep{cer96}.
Nevertheless, radiative transfer calculations appear to indicate that 
HCN radiative excitation through absorption of infrared photons emitted 
by warm dust dominates over collisional excitation
\citep{gon99a,din00}, and in such a case the observed HCN emission
would not involve any further cooling.

\begin{figure}
\epsscale{0.8}
\plotone{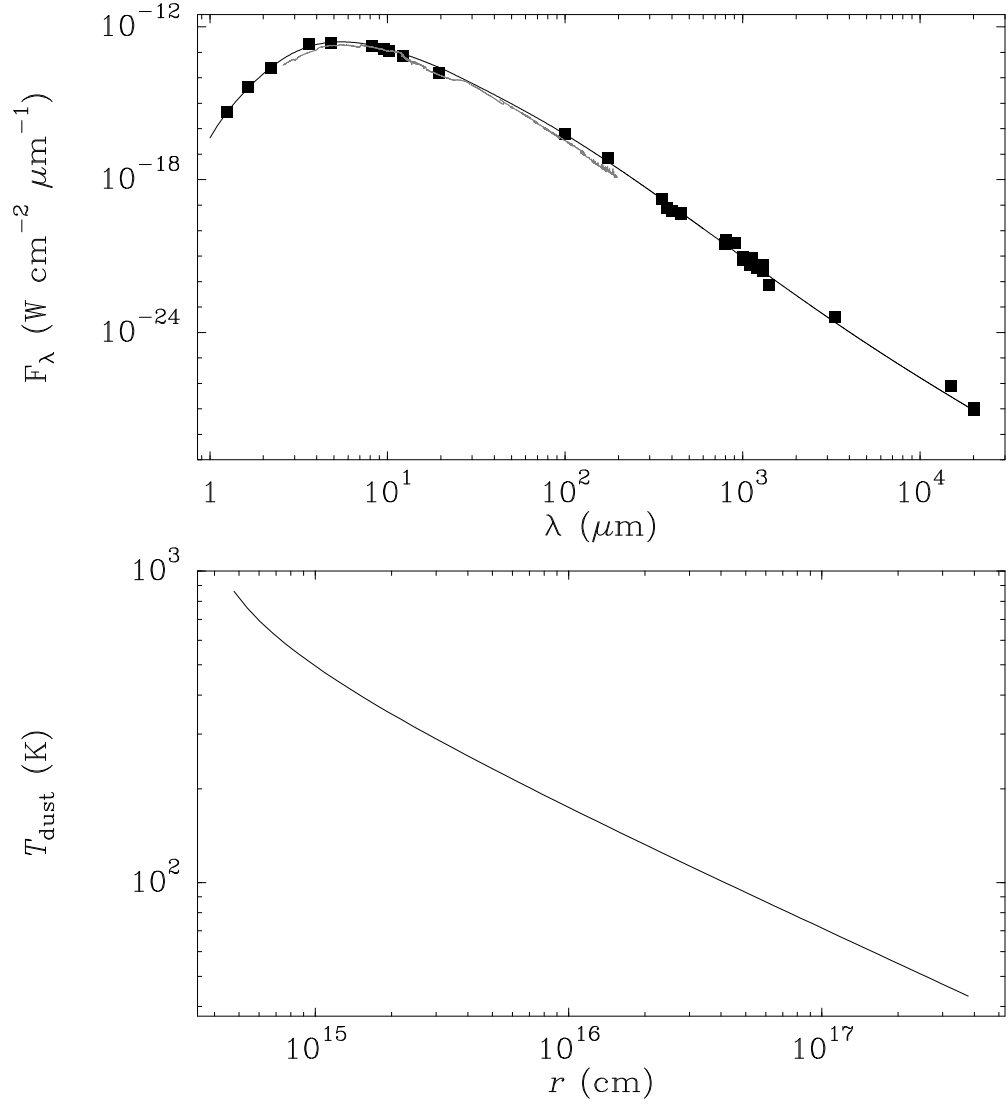}
\caption{{\em Top}: Fit to the continuum emission from IRC+10216. 
Data points are taken from \cite{bag95}, and the grey line shows
the ISO data. {\em Bottom}: Dust temperature profile. 
\label{fig:cont}}
\end{figure}

It is shown below that the continuum emission provides the dominant 
excitation source for all \hdo\ lines of interest, mainly through 
absorption of 6 $\mu$m photons in the $\nu_2$ band and subsequent decay, 
and so a model for the dust emission from IRC+10216 is required. 
We have used a radiative transfer code to compute, from the balance 
between heating and cooling at any radial
position in the envelope, the dust temperature profile \citep{gon99b}.
Our model simulates the dust properties by making use of an empirical 
spectral index $\beta$. The inferred dust parameters are listed in 
Table~\ref{tab:envel}, and the goodness of the fit, shown in
Fig.~\ref{fig:cont}, ensures that the continuum radiation density is 
well reproduced at least at $r\gtrsim2\times10^{15}$ cm, where it varies 
as $\sim r^{-2}$.

The radiative transfer code used to determine the \hdo\ excitation and 
emission has been described in \cite{gon97,gon99b}. The calculations 
assume spherical symmetry, and the envelope is divided into 60 spherical 
shells to account for the variations in the physical and chemical conditions 
with distance to the star. The dust and \hdo\ are assumed to be 
coexistent within any shell, and the non-local radiative transfer is 
simulated through rays that cross the source. The code first computes 
the \hdo\ statistical equilibrium populations at any radial position in 
the envelope with the use of an iterative procedure based on an 
approximate Newton-Raphson operator \citep{sch86}. 
A maximum relative variation of populations of $10^{-3}$ was established
as the convergence criterion; a much tighter criterion was found
in some test models to change line fluxes by at most 0.2\%. Once 
convergence is achieved, the resulting line profiles and fluxes are 
calculated. We report the expected intensities in units of main-beam 
temperature ($T_{MB}$), with a half-power 
beam width of HPBW=240$''$ to simulate SWAS measurements, and HPBW ranging 
from $12\arcsdot5$ (1703-1910 GHz) to $41''$ (480-640 GHz) for HSO-HIFI 
observations (see Table~\ref{tab:h2otrans}, below).

Spectroscopic data for \hdo\ were taken from the Hitran 2004 database 
\citep{rot05}. We included the ground and the $\nu_2=1$
and $\nu_3=1$ excited vibrational states, but the $\nu_3=1$ state was found
to have a negligible effect on the emission from the ground state lines.
The $\nu_1=1$ state was ignored because the $\nu_1$ band is much
weaker than the $\nu_3$ band, and more excited vibrational levels are also
ignorable because they are weakly connected to the ground vibrational
state and/or the continuum emission from IRC+10216 weakens in the
near infrared.
Up to 40 rotational levels per vibrational state for both ortho-\hdo\ and
para-\hdo\ were taken into account, but the code allows this number 
to be decreased in external regions with low excitation, with the 
aim of ensuring and accelerating convergence. 

The calculations simulate both collisional excitation among rotational 
levels of the ground vibrational state, and radiative excitation in all 
transitions. Vibrational excitation through collisions is neglected. 
Collisional rates among the 10 lowest rotational levels for $T_k\le140$ K 
are taken from \cite*{phi96}. We extrapolate these rates to higher $T_k$ 
by multiplying the excitation rates for $T_k=140$ K by 
$\exp\{\Delta E(1/140-1/T_k)\}$, where $\Delta E (K)=E_{up}-E_{low}$ is the 
energy difference between the upper and lower levels. 
The application of this strategy appears to be justified, since the
use of the \cite{phi96} collisional rates at $T_k=140$ K, together with the 
quoted extrapolation, yield collisional rates for $T_k=20-120$ K that 
reproduce the values given by \cite{phi96} within a factor of 1.6.
Also, some estimate of
the collisional rates among levels not considered by \cite{phi96} is 
required to simulate the collisional excitation in the innermost regions 
of the envelope. With this aim, we have found that if the \hdo-He collisional 
excitation rates for the 10 lowest rotational levels given by 
\cite*{gre93}, are multiplied by a global 
factor of 5, {\em most} of the \hdo--H$_2$ $J=1$ rates given by \cite{phi96} 
are reproduced within a factor of 2.5; similarly, {\em most}
of the \hdo--H$_2$ $J=0$ rates given by \cite{phi96} are reproduced
within the same factor if the Green et al. rates are multiplied by
a factor of 1.35, the difference in reduced mass when H$_2$ is the
collision partner rather than He. After applying these
corrections to the \cite{gre93} coefficients, some individual rates are
still strongly underestimated, by 1-2 orders of magnitude, relative to the 
\cite{phi96} values, specially in the case of collisions between 
para-\hdo\ $0_{00}$ and H$_2$ $J=1$ \citep[see discussion in][]{phi96}.
Nevertheless, the collisional excitation of a high-energy level in
high-density and high-temperature regions is the result of collisional 
pumping from 
various other levels, and so individual discrepancies are to some extent 
diluted. In summary, we have used the \cite{gre93} excitation rates, 
multiplied by the factors quoted above, to simulate the collisional 
excitation among levels not considered in \cite{phi96}. The adopted 
extrapolation does not critically affect results because, as shown below, 
radiative excitation dominates over collisional excitation in all lines 
of interest.

We have generated a set of models that differ only in the inner radius
of the \hdo\ shell, $R_{int}$, and in the \hdo\ abundance, $X$(\hdo).
The different values of $R_{int}$ account for different spatial 
distributions of \hdo\ in the envelope of IRC+10216, and $X$(\hdo) is
derived in such a way that the predicted flux of the o-\hdo\ \t110101\ is 
$\approx10^{-20}$ W cm$^{-2}$ as observed by SWAS \citep{mel01}. 
The ortho-to-para \hdo\ abundance ratio is assumed to be 3:1. In all
models, we have adopted an external radius of the \hdo\ shell 
$R_{out}=4\times10^{17}$ cm, where \hdo\ is assumed to be photodissociated.
This value of $R_{out}$ is similar to that obtained by \cite{wil04}
and AC06 through photodissociation models of the external shells of the
IRC+10216 circumstellar envelope.
The photodissociation radius depends on several parameters and 
could be significantly lower than the quoted value \citep{for03,net87}, 
but test models indicate that line flux ratios for 
$R_{int}\lesssim8\times10^{15}$ cm vary by less than 20\% if $R_{out}$ 
is decreased to $5\times10^{16}$ cm.
For high values of $R_{int}$, we give in section   
\ref{sec:radassoc} an analytic expression for the expected flux of the
o-\hdo\ \t110101\ line as a function of both $R_{int}$ and $R_{out}$,
showing that results are not very sensitive to $R_{out}$ as long as
$R_{out}/R_{int}\gg1$.

We have selected three of those models, called $A$, $B$, and $C$, that 
correspond to the proposed formation mechanisms described in 
section \ref{sec:intro}. The adopted values of $R_{int}$ and the
inferred ortho-\hdo\ abundance, $X$(o-\hdo), are listed in 
Table~\ref{tab:h2omodels}. Model $A$ simulates \hdo\ originating 
in the innermost regions of the envelope. 
Model $B$ applies to both the cometary hypothesis and the Fischer-Tropsch 
catalysis mechanism. Vaporization of icy bodies predicts that \hdo\
is released at a radius of $(1-5)\times10^{15}$ cm, depending 
on the evolutionary stage of the star after the onset of the TP-AGB phase  
\citep{for01}. The formation of \hdo\ due to Fischer-Tropsch catalysis on 
the surfaces of iron grains predicts that the \hdo\ abundance attains its 
maximum value at $(1.5-2)\times10^{15}$ cm \citep{wil04}. Therefore, these
two models predict similar \hdo\ spatial distributions, although the
Fischer-Tropsch mechanism appears to be more restrictive.
Model $C$ assumes \hdo\ to be generated as a consequence of 
photodissociation in the outermost regions of the envelope (AC06).

\section{Model results} \label{sec:results}

\subsection{The H$_2$O $1_{10}-1_{01}$ line in IRC+10216}
\label{sec:110101}

\begin{figure}
\epsscale{1}
\plotone{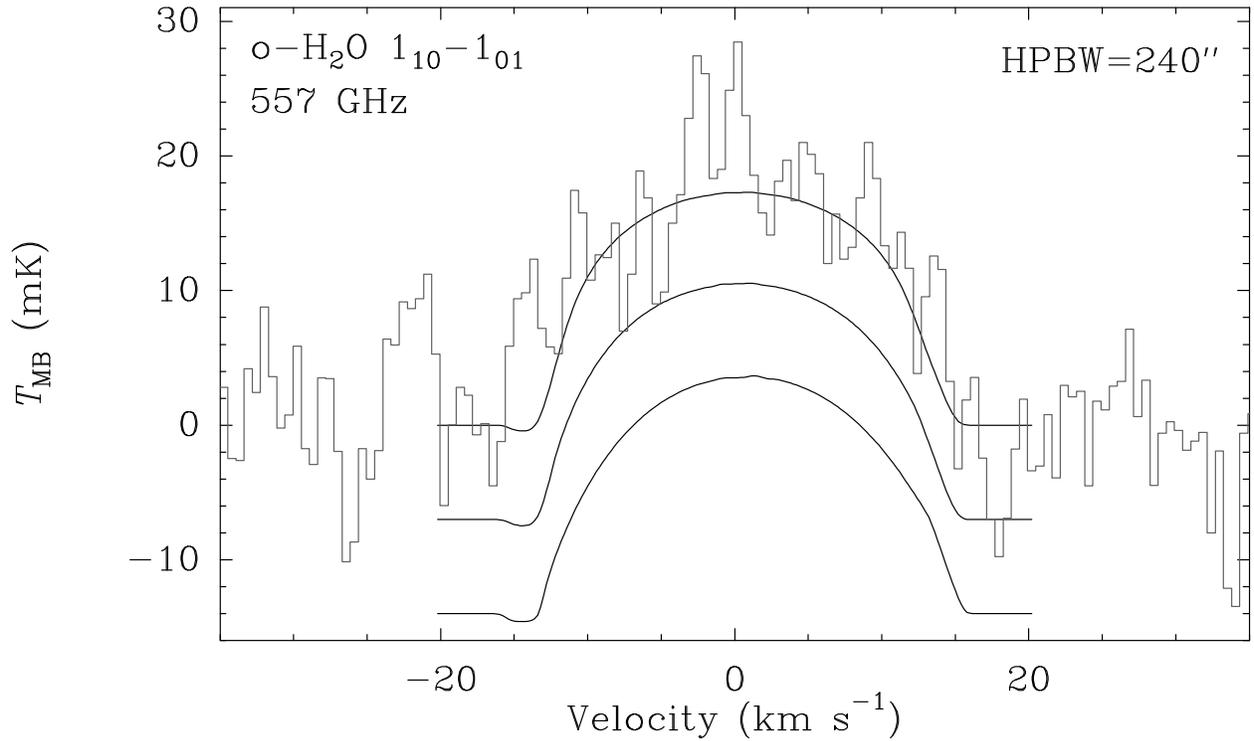}
\caption{Comparison between the SWAS \t110101\ o-\hdo\ spectrum
\citep[from][]{mel01} and the line profile obtained for models
$A$ (upper), $B$ (middle), and $C$ (lower), with the SWAS beamsize.
The spectra of models $B$ and $C$ have been shifted vertically for clarity.
The required abundances of o-\hdo\ are given in Table~\ref{tab:h2omodels}. 
\label{fig:prof557ghz}}
\end{figure}

\begin{figure}
\epsscale{1}
\plotone{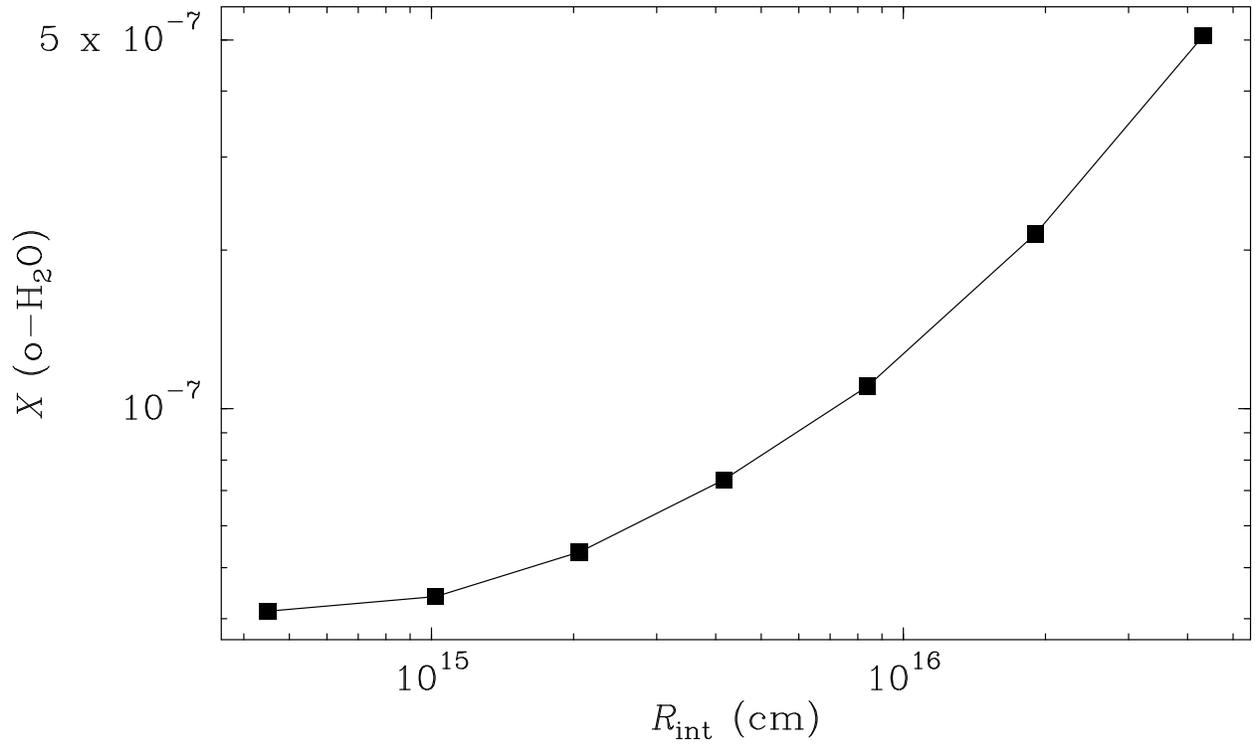}
\caption{Ortho-\hdo\ abundance, relative to H$_2$, required to
account for the observed 557 GHz \hdo\ emission in IRC+10216,
as a function of the assumed inner radius of the \hdo\ shell.
\label{fig:xoh2orint}}
\end{figure}

\begin{figure}
\epsscale{1}
\plotone{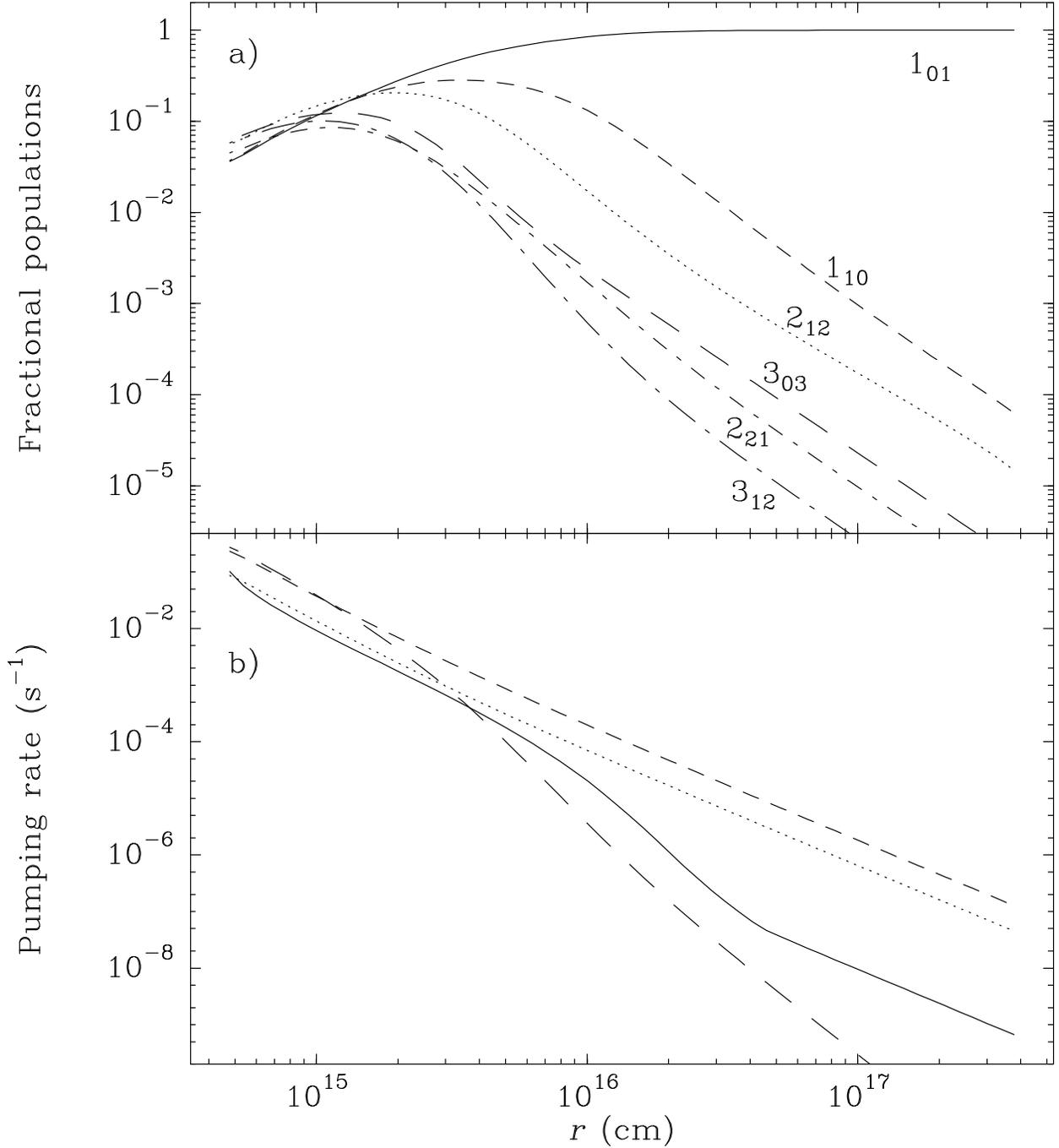}
\caption{
a) Fractional populations of the 6 lowest energy levels of 
ortho-\hdo\ in model $A$;
b) collisional and radiative pumping rates, per molecule
in the $1_{01}$ level, for the $1_{10}$ level in model $A$. 
Solid line: collisional pumping, which is computed from the 
$1_{01}\rightarrow1_{10}$ collisional rate; dashed line: 
radiative rate through the path 
$1_{01}\rightarrow\nu_2\,1_{10}\rightarrow2_{21}\rightarrow1_{10}$;
dotted line: radiative rate through the path 
$1_{01}\rightarrow\nu_2\,2_{12}\rightarrow2_{21}\rightarrow1_{10}$;
long-dashed line: radiative rate through the path 
$2_{12}\rightarrow\nu_2\,1_{01}\rightarrow1_{10}$.
\label{fig:pump557ghz}}
\end{figure}

Figure~\ref{fig:prof557ghz} compares the SWAS \t110101\ 
ortho-\hdo\ continuum-subtracted spectrum \citep[from][]{mel01} with the
line profiles obtained for models $A$, $B$, and $C$, with the SWAS
beamsize. All models yield similar line profiles because in all
three cases the emission region is small compared to the SWAS $4'$
beam. The line shapes obtained with the smaller ODIN $2'$ beam (not shown)
are also similar because, even in model $C$, 70\% of the emission 
arises within a spherical shell of diameter $1\arcmdot5$
($2.3\times10^{17}$ cm), which is
significantly smaller than the ODIN beamsize. The o-\hdo\ abundance 
(relative to H$_2$) required to account for the observed flux, is plotted
in Fig.~\ref{fig:xoh2orint} as a function of $R_{int}$; it increases
slowly for low $R_{int}$ because the innermost regions of the envelope
contribute little to the emission at 557 GHz. For $R_{int}=2\times10^{15}$ cm
(model $B$), the required ortho-\hdo\ abundance, 
$X$(o-\hdo)$=5.4\times10^{-8}$ (Table~\ref{tab:h2omodels}),
is one order of magnitude lower than that reported previously by
\cite{mel01} and \cite{has06}.
As first noted by AC06 for the case of high $R_{int}$,
this result is a consequence of the pumping of the $\nu_2=1$ vibrational
state by dust emission at 6 $\mu$m and the subsequent decay to the ground 
state, which dominates the excitation of \hdo\ in our models. 
In the case of the o-\hdo\ 557 GHz line, the importance of mid-IR pumping
can be shown as follows.
The radiative path that dominates the pump of the $1_{10}$ rotational level is
$1_{01}\rightarrow\nu_2\,1_{10}\rightarrow2_{21}\rightarrow1_{10}$. The
associated pumping rate per molecule in the $1_{01}$ level is given 
by $\Gamma_r=B_{lu}\langle J\rangle\,\eta_1\eta_2$, where $B_{lu}$ is
the Einstein coefficient for absorption of radiation in the
$1_{01}\rightarrow\nu_2\,1_{10}$ line, $\langle J\rangle$ is the continuum 
intensity at 6.2 $\mu$m averaged over all angles and integrated over 
the line profile, $\eta_1$ is the fraction of molecules in $\nu_2\,1_{10}$ 
that decay to $2_{21}$, and $\eta_2$ is the fraction of molecules in 
$2_{21}$ that decay to $1_{10}$. Ignoring line opacity effects, and 
estimating $\langle J\rangle$ as
\begin{equation}
\langle J\rangle = \frac{D^2}{4\pi\,r^2} F_{6.2\,{\rm \mu m}}, 
\label{eqj}
\end{equation}
where $F_{6.2\,{\rm \mu m}}=2.5\times10^{-13}$ W cm$^{-2}$ $\mu$m$^{-1}$  
is the observed 6.2 $\mu$m continuum flux, and $r$ is the distance from
  the star, one obtains
\begin{equation}
\Gamma_r\approx\frac{0.02}{r_{15}^{2}} \,\, {\rm s^{-1}}, 
\label{gammar}
\end{equation}
where $r_{15}=r/(10^{15} \, {\rm cm})$.
The ro-vibrational lines are indeed optically thin in all our models.
The $1_{01}\rightarrow\nu_2\,1_{10}$ and $1_{01}\rightarrow\nu_2\,2_{12}$
have similar opacities; for $R_{int}\gtrsim2\times10^{16}$ cm, essentially
all o-\hdo\ molecules are in the ground $1_{01}$ level
(Fig.~\ref{fig:pump557ghz}a) and the radial
opacities of the quoted lines are given by
\begin{eqnarray}
\tau_r & \approx & 3\times10^{-2} \times 
\left(\frac{0.65 \, {\rm km \, s^{-1}}}{\sigma_v}
  \right) \times \left(\frac{X({\rm o-H_2O})}{5\times10^{-7}}\right) \nonumber
  \\
& \times & \left(\frac{R_{out}}{R_{int}}-1 \right)
\times \left(\frac{4\times10^{17} \, {\rm cm}}{R_{out}}\right),
\label{eq:taur}
\end{eqnarray}
which gives $\tau_r\approx0.25$ for model $C$. The fraction of o-\hdo\
molecules in excited energy levels is 20--30\% at $r\sim8\times10^{15}$ cm
and increases sharply with diminishing $r$ (Fig.~\ref{fig:pump557ghz}a);
therefore, eq. (\ref{eq:taur}) overestimates the opacities for
$R_{int}\lesssim8\times10^{15}$ cm. Our models yield maximum values of 
$\tau_r=0.30$ for the case of $R_{int}=4\times10^{15}$ cm. All other
ro-vibrational o-\hdo\ transitions have excited lower levels and hence 
lower opacities.

On the other hand, the collisional pumping rate from $1_{01}$ to $1_{10}$
per molecule in the $1_{01}$ level is $\Gamma_c=n({\rm H}_2)\,C_{lu}(T_k)$, 
where $C_{lu}$ is the collisional rate from the $1_{01}$ to the $1_{10}$ 
level, and can be written as
\begin{eqnarray}
\Gamma_c & \approx & \frac{0.01}{r_{15}^{2}}  \times
\left(\frac{C_{lu}(300 {\rm K})}{2.3\times10^{-10}\,\,{\rm cm^{3}\,s^{-1}}}
\right) \nonumber \\
& \times & \exp\left\{-120\left(\frac{1}{T_k(r)}-\frac{1}{300}\right)\right\}
\,\, {\rm s^{-1}},
\label{gammac}
\end{eqnarray}
where the exponential factor is an approximate fit to the variation
of $C_{lu}$ with $T_k$, valid up to $r\approx3\times10^{16}$ cm.
Equations (\ref{gammar}) and (\ref{gammac}) show that even at $r_{15}=1$,
where $T_k\approx300$ K, the radiative-to-collisional pumping rate ratio
is $\sim2$, and strongly increases with increasing $r$ as a consequence of 
the diminishing $T_k$. Figure~\ref{fig:pump557ghz}b shows the actual values 
of $\Gamma_r$ (dashed line) and $\Gamma_c$ (solid line) given by the code 
in model $A$. At $r_{15}=1$, $\Gamma_r$ is a factor of $\approx2$ higher 
than predicted by eq. (\ref{gammar}) because dust and gas are mixed and 
$\langle J\rangle$ is higher than estimated above. The contribution to
the excitation of other radiative paths, like 
$1_{01}\rightarrow\nu_2\,2_{12}\rightarrow2_{21}\rightarrow1_{10}$
and $2_{12}\rightarrow\nu_2\,1_{01}\rightarrow1_{10}$, accounts
for the relatively low o-\hdo\ abundance required to reproduce the
observed 557 GHz flux.

As pointed out in section \ref{sec:model}, the inclusion of the 
$\nu_3=1$ state has
negligible effect on the line fluxes from the ground vibrational state,
which can be shown as follows. The strongest line connecting the
$1_{01}$ level and the $\nu_3=1$ state is the
$1_{01} \rightarrow \nu_3\, 2_{02}$ transition at $\lambda=2.63$ $\mu$m, 
with a pumping rate per molecule in the $1_{01}$ level given by
$\Gamma_{r3}\propto\lambda^5\, g_{up}\, A_{ul}\, F_{2.6\,{\rm \mu m}}$;
here $g_{up}$ is the degeneracy of the upper level of the transition,
$A_{ul}\approx33$ s$^{-1}$ is the Einstein coefficient for spontaneous 
emission, and $F_{2.6\,{\rm \mu m}}\approx0.6\times10^{-13}$ W
cm$^{-2}$ $\mu$m$^{-1}$ is the observed 2.6 $\mu$m continuum flux. 
A similar expression applies to the pumping rate in the
$1_{01} \rightarrow \nu_2 \, 1_{10}$ transition at 6.2 $\mu$m 
($A_{ul}\approx10$ s$^{-1}$), so that the 
ratio of the pumping rates from the $1_{01}$ level into the $\nu_2 \, 1_{10}$
and $\nu_3\, 2_{02}$ levels is found to be $\approx55$.

\begin{figure}
\epsscale{1}
\plotone{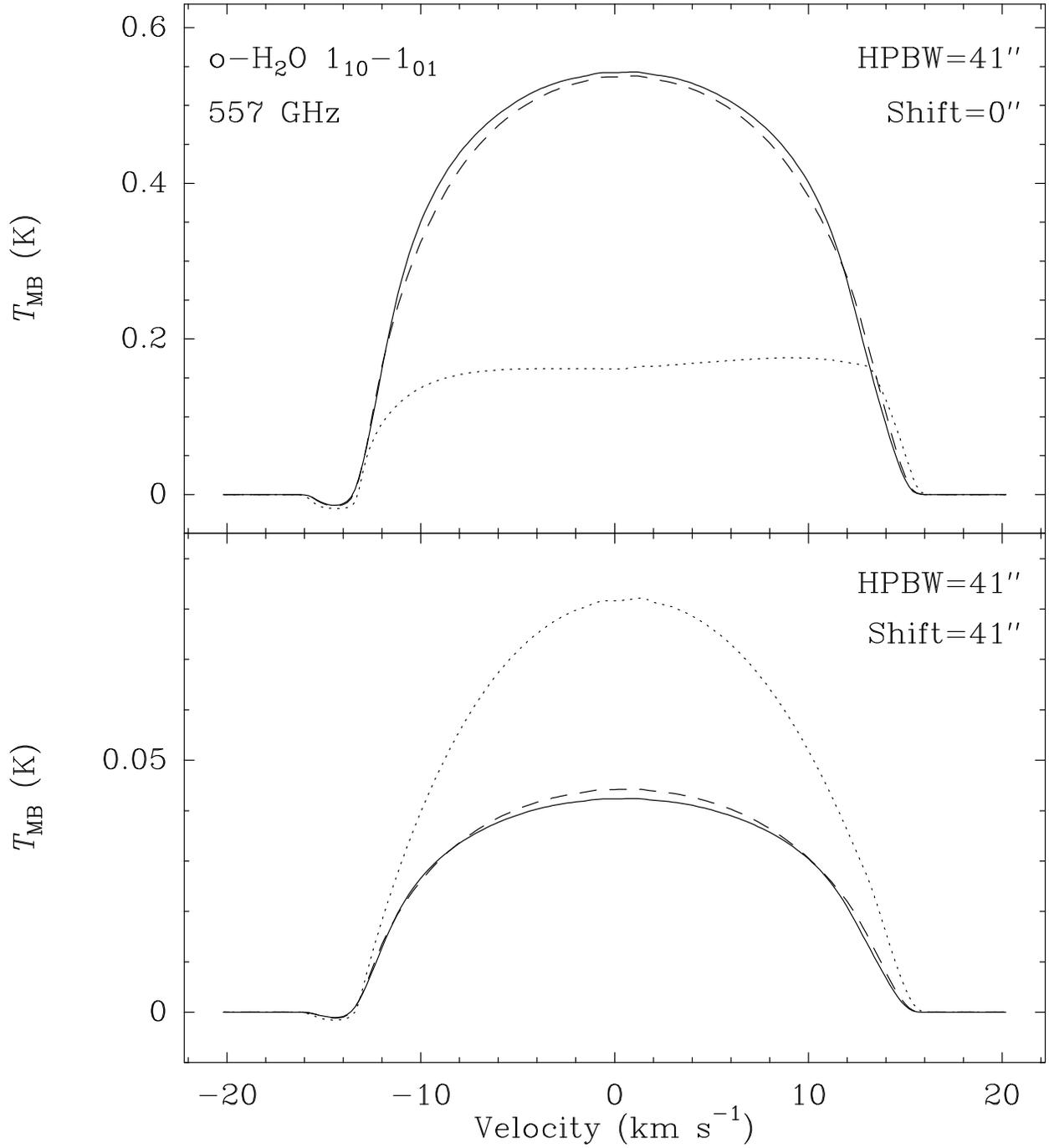}
\caption{Predicted line profile of the o-\hdo\ \t110101\ line with the 
HSO-HIFI beam, and the telescope axis pointing toward the source center
({\em upper}) and toward a position shifted $41''$ from the source center
({\em lower}). Model $A$: solid line;   
$B$: dashed line; $C$: dotted line. 
\label{fig:prof557ghzhso}}
\end{figure}

Figure~\ref{fig:prof557ghzhso} shows the predicted 557 GHz line profiles
as observed with HSO (HPBW=$41''$). Results are given for the telescope
pointing toward the source center, and shifted $41''$ from the source
center. In models $A$ and $B$, the contribution to the total emission 
of the outermost regions ($r>4\times10^{16}$ cm, or $16''$) to the 557 
GHz emission is low, and so the line shapes and relative intensity toward 
the offset position are typical of an optically thick, spatially unresolved 
line. Model $C$ shows, on the contrary, 
a nearly flat line profile with relatively weak emission toward the source 
center because the source is partially resolved; in this case the intensity 
toward the offset position is the strongest of the three models. The
predicted line flux and shape within the HSO beam is somewhat dependent
on the assumed value of $R_{out}$, but significant loss of line flux in
the wings of the beam is expected in model $C$ as long as the source size, 
$2\times R_{out}$, is larger than the beam size at 557 GHz, $10^{17}$ cm 
(i.e. $R_{out}/R_{int}>1.25$).

\subsection{Predictions for other lines in IRC+10216} \label{sec:predictions}

\begin{figure}
\epsscale{1}
\plotone{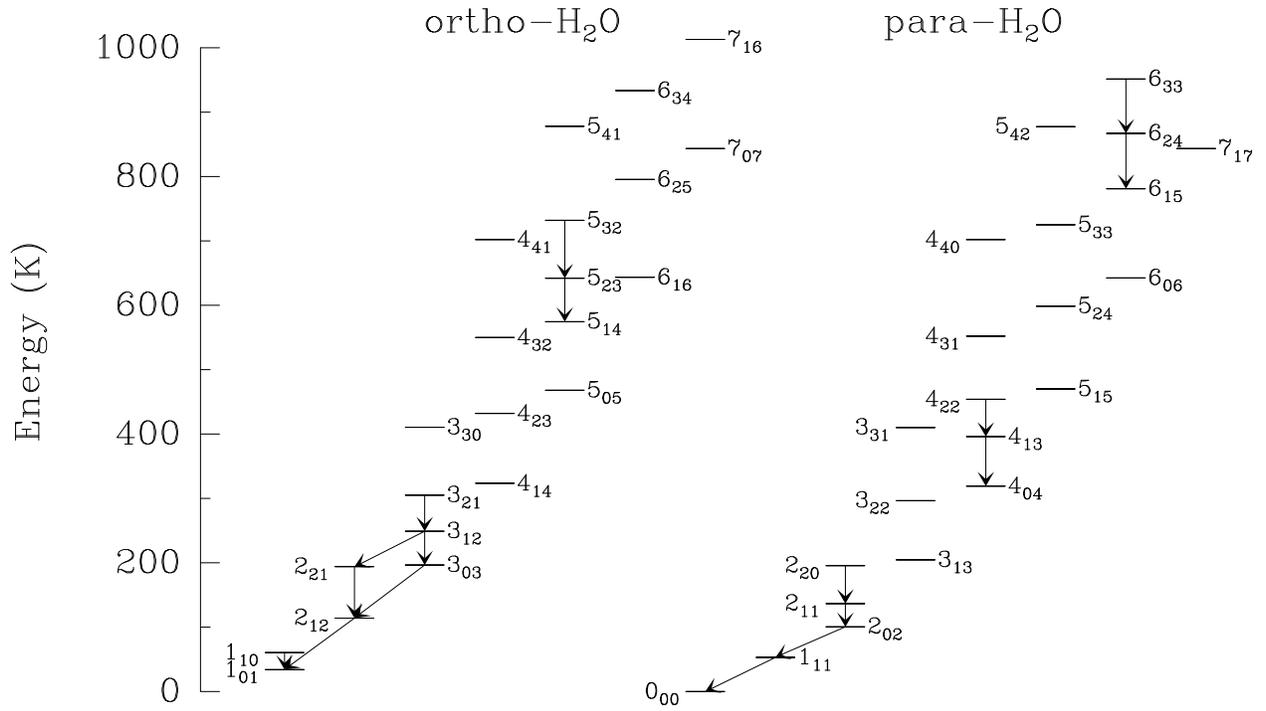}
\caption{Energy level diagram of ortho and para-\hdo. Arrows indicate
the transitions whose predicted line fluxes and profiles are shown in 
Figs. \ref{fig:flujosrint2} and \ref{fig:lines}.  
\label{fig:diagram}}
\end{figure}

\begin{figure}
\epsscale{1}
\plotone{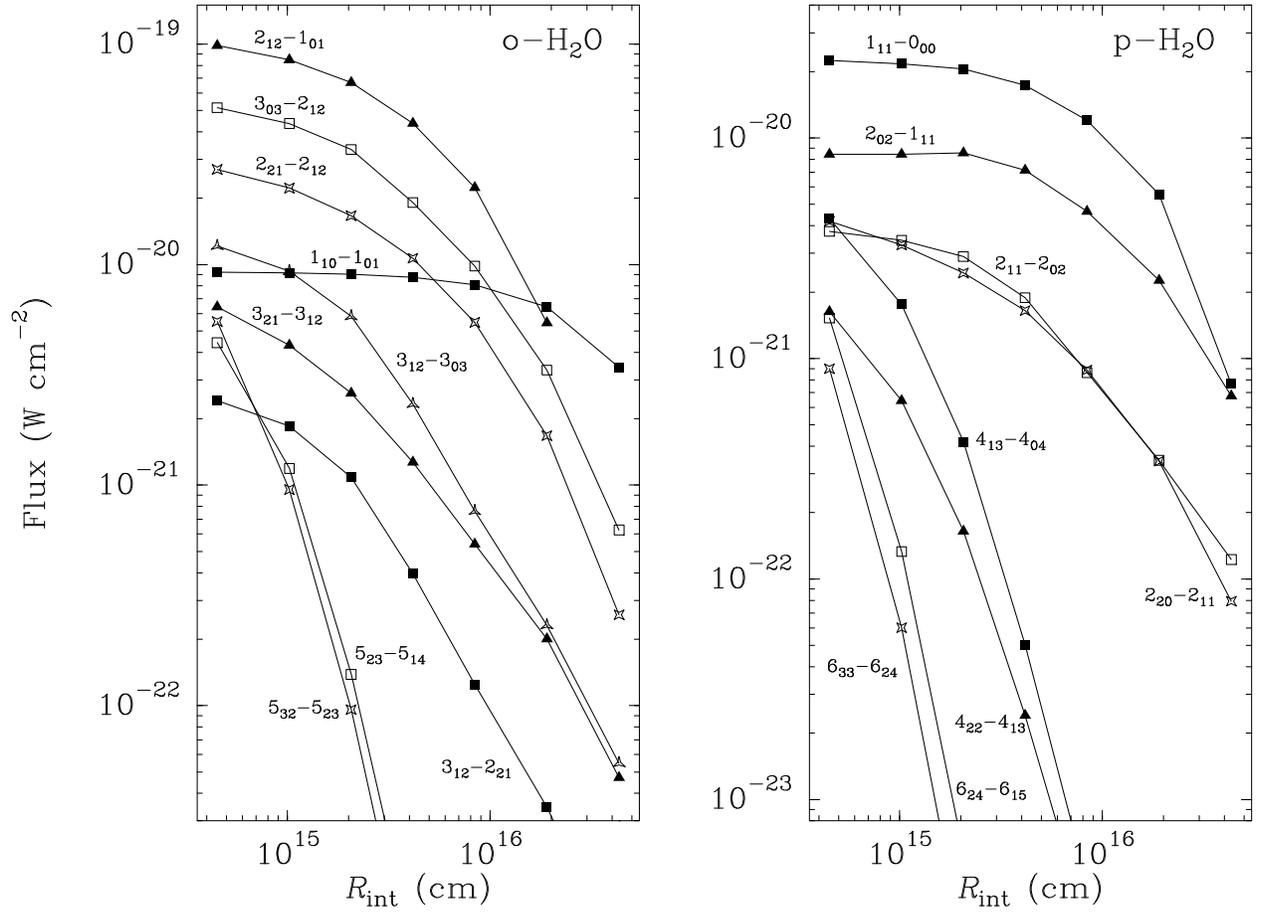}
\caption{Predicted \hdo\ line fluxes observable by HSO-HIFI as a 
function of the inner radius of the \hdo\ shell. Adopted beamwidths
are listed in Table \ref{tab:h2otrans}. 
\label{fig:flujosrint2}}
\end{figure}

\begin{figure}
\epsscale{0.8}
\plotone{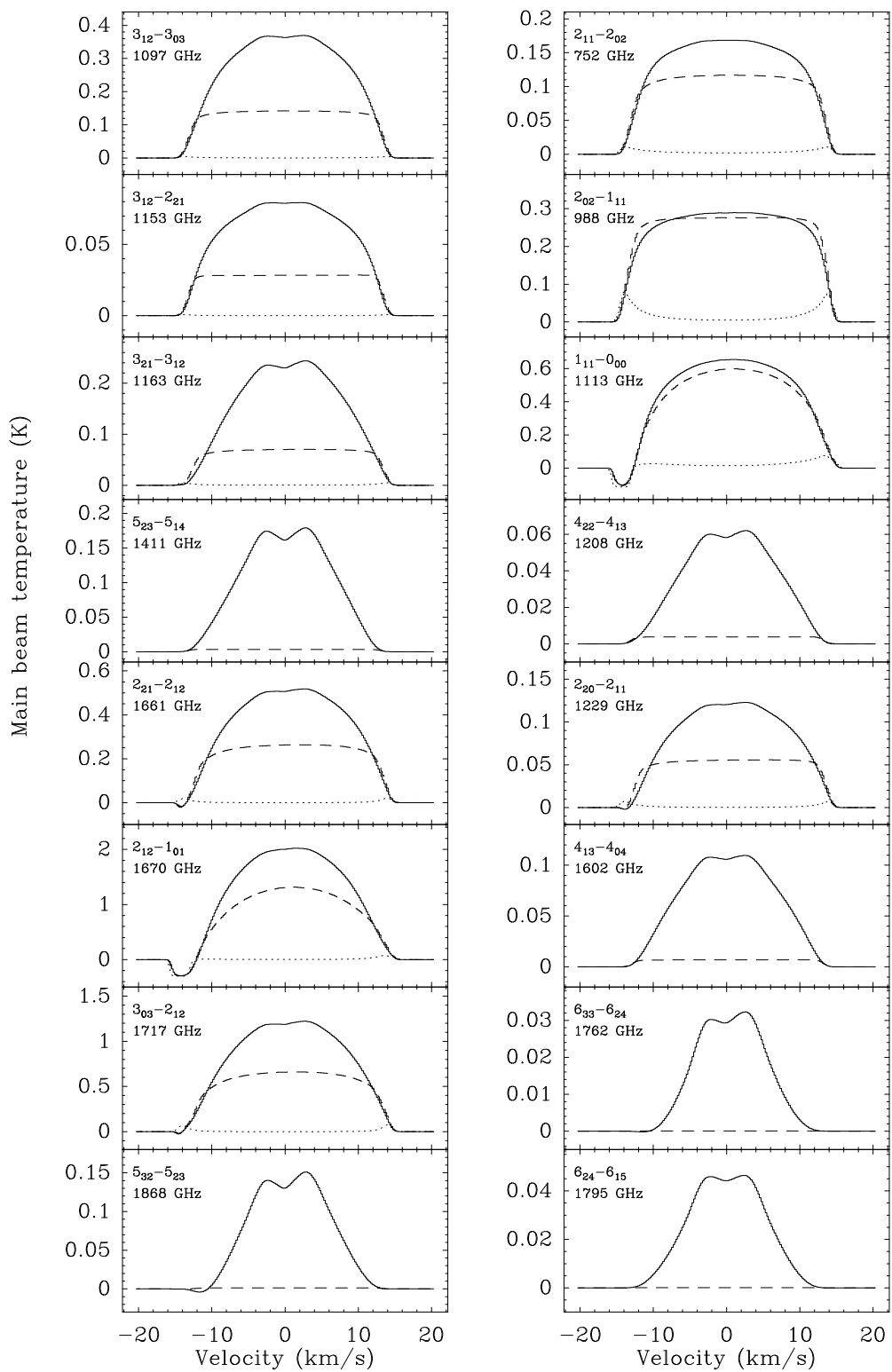}
\caption{Predicted \hdo\ line profiles observable by 
HSO-HIFI, for models $A$ (solid), $B$ (dashed), and $C$ (dotted). 
Adopted beamwidths are listed in Table \ref{tab:h2otrans}. 
\label{fig:lines}}
\end{figure}

The models presented below, showing predictions for \hdo\ lines other
than the 557 GHz one, are the same as those already described in 
the previous sections, with values for the water vapor abundance such
that the o-\hdo\ 557 GHz line flux observed by SWAS is reproduced
for all assumed $R_{int}$ (Fig.~\ref{fig:xoh2orint}).
Figure~\ref{fig:diagram} shows the energy level diagram of ortho and
para-\hdo, and indicates the strongest lines 
($\Delta K_{-1}=\pm1$, $\Delta K_{+1}=\pm1$) that lie within the
HSO-HIFI allowed frequency range.
With reference to the energy of the upper level, the lines may be 
classified into three groups: $(i)$ 8 low-excitation
lines, with $E_{up}\lesssim200$ K; $(ii)$ 3 mid-excitation lines, with 
$200 \, {\rm K}<E_{up}\lesssim300$ K; $(iii)$ 6 high-excitation lines, with
$E_{up}>300$ K ($J_{up}\ge4$). 
Figure~\ref{fig:flujosrint2} plots the
predicted line fluxes, as observed with the HSO-HIFI beam, versus the inner 
radius of the \hdo\ shell, and Fig.~\ref{fig:lines} shows the expected 
line profiles for models $A$, $B$, and $C$. Line frequencies and HSO-HIFI
beamwidths adopted in these calculations are listed in 
Table~\ref{tab:h2otrans}.

The overall excitation is, as in the case of the 557 GHz line, dominated
by the pumping of the $\nu_2$ state through absorption of 6 $\mu$m 
dust-emitted photons. Nevertheless, absorption of continuum photons
in far-infrared pure-rotational transitions also plays an important role 
in the excitation of some mid- and high-excitation lines. For example, 
absorption of 45 $\mu$m and 58 $\mu$m photons in the \t523414\ and \t422313\ 
lines, contributes significantly to the excitation of the \t523514\ and 
\t422413\ lines, respectively. On the other hand, we find that the effect 
of collisions is ignorable in all our models. We have run model $A$ by 
quenching off all collisions among rotational levels, and found that the 
line fluxes vary by less than 4\%. Collisional rates much higher than 
estimated in our extrapolation would be required for collisions to compete 
with the radiation field. We conclude that the excitation of \hdo\ is 
dominated by the radiation field in all lines and models and our results 
are dependent on neither the H$_2$ density nor the $T_k$ profiles, but 
rely on the adequate simulation of the radiation field generated by dust.

Line fluxes decrease with increasing $R_{int}$ because 
$(i)$ owing to the decrease of the overall \hdo\ excitation with
increasing distance from the star (Fig.~\ref{fig:pump557ghz}a), 
the emissivity of all lines decreases with increasing
$R_{int}$ more steeply than that of the lowest-lying \t110101\
transition, and since the SWAS flux of the latter line is 
kept constant for all $R_{int}$, all other line fluxes decrease;
$(ii)$ for high $R_{int}$ and low-excitation lines, the emission becomes
spatially resolved, and hence line flux is lost beyond the wings of the
HSO-HIFI beam. The latter effect becomes quite pronounced for 
high-frequency lines, and explains why the \t212101\ line 
(HPBW=$14\arcsdot5$, corresponding to a radius of 
$\approx2\times10^{16}$ cm) is predicted to be in absorption against the 
179.5 $\mu$m continuum emission for $R_{int}\gtrsim4\times10^{16}$ cm 
(Fig.~\ref{fig:lines}).

Lines belonging to different groups trace different ranges of $R_{int}$. 
The low-excitation lines will allow us to check whether \hdo\ is
formed as a result of photodissociation products.
If \hdo\ is formed in the outermost layers of the envelope, with
$R_{int}\gtrsim2\times10^{16}$ cm, the 557 GHz line is expected to be the
strongest of all lines. Most lines would be undetectable in such 
a case, given the expected sensitivity of HSO-HIFI 
(Table~\ref{tab:h2otrans}).

The observation of the high and mid-excitation lines would check whether 
\hdo\ is formed in the innermost layers of the envelope. The fluxes 
predicted for the high-excitation lines decrease by one order of 
magnitude or more for 
$4.5\times10^{14}\,{\rm cm}\le R_{int}\leq2\times10^{15}$ cm;
however, these high-frequency lines are hardly detectable with the expected 
sensitivity of HSO-HIFI. Also, the predicted fluxes
are somewhat uncertain because they depend on the details of the radiation 
field in the innermost regions of the envelope, which in
turn depend on the distribution and emissivity of dust in these regions.
Nevertheless, some mid-excitation lines are still sensitive to low values
of $R_{int}$, as, for example, the \t312303\ and \t321312
transitions. These lines, as well as 
other low-excitation lines, will also probe intermediate values of 
$R_{int}$. The fluxes of the \t312303\ and \t312221\ lines vary by 
a factor of $\approx2.5$ from $R_{int}=2\times10^{15}$ cm to 
$R_{int}=4\times10^{15}$ cm, and the flux of the \t303212\ line varies 
by a factor of $\approx2$ in the same $R_{int}$ interval.

The expected line profiles (Fig.~\ref{fig:lines}) display a variety of 
shapes, from parabolic (optically-thick, spatially-unresolved lines)
to triangular (high-excitation lines), double-peaked 
(spatially-resolved lines), top-hat (optically-thin, 
spatially-unresolved lines) and P-Cygni (spatially-resolved, ground-state 
lines) profiles. The greater uncertainty concerns the predicted shapes of 
the high-excitation lines, as they are formed in the acceleration region.

\subsection{HSO observations of C-rich AGB stars other than IRC+10216}
\label{sec:crich}

\begin{figure}
\epsscale{0.8}
\plotone{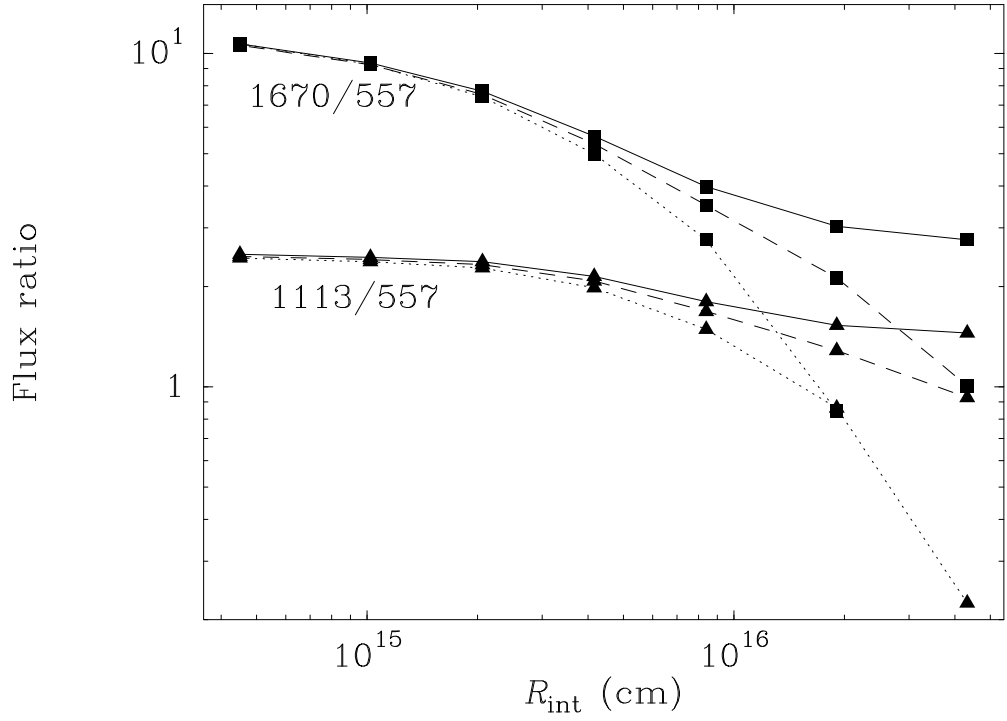}
\caption{Calculated $F$(\t212101)/$F$(\t110101) (labelled 1670/557,
squares) and $F$(\t111000)/$F$(\t110101) (labelled 1113/557, triangles)
vs. $R_{int}$ for the IRC+10216 model source, but located at a distance
of $D>2$ kpc (solid lines), $D=0.5$ kpc (dashed lines), and 
$D=0.17$ kpc (dotted lines).
\label{fig:radiosrint}}
\end{figure}

In this section we explore the prospects for detecting any of the low-lying
ground vibrational state \hdo\ lines, \t110101\ at 557 GHz, \t111000\ 
at 1113 GHz, and \t212101\ at 1670 GHz, in C-rich AGB stars other than 
IRC+10216. Figure \ref{fig:radiosrint} shows that, for a source similar to
IRC+10216, the line flux ratios depend for high $R_{int}$ on the
distance to the source, as a consequence of the spatially extended emission 
and of the different HSO/HIFI beams available for the three lines
(Table~\ref{tab:h2otrans}). For $R_{int}\approx4\times10^{16}$ cm,
the 1670 GHz line will only be spatially unresolved if the source is 
located at more than 2 kpc. We will restrict in the following to models 
with $R_{int}=2\times10^{15}$ cm, which predict spatially unresolved 
emission in all lines at essentially any distance.
This value of $R_{int}$ corresponds to model $B$ in the previous sections,
which assumes \hdo\ to be formed or released at intermediate radii, and
whose results for the low-lying lines considered here are also very similar 
to those obtained for the lowest $R_{int}$ (model $A$, see
Table~\ref{tab:h2omodels}).

\begin{figure}
\epsscale{0.6}
\plotone{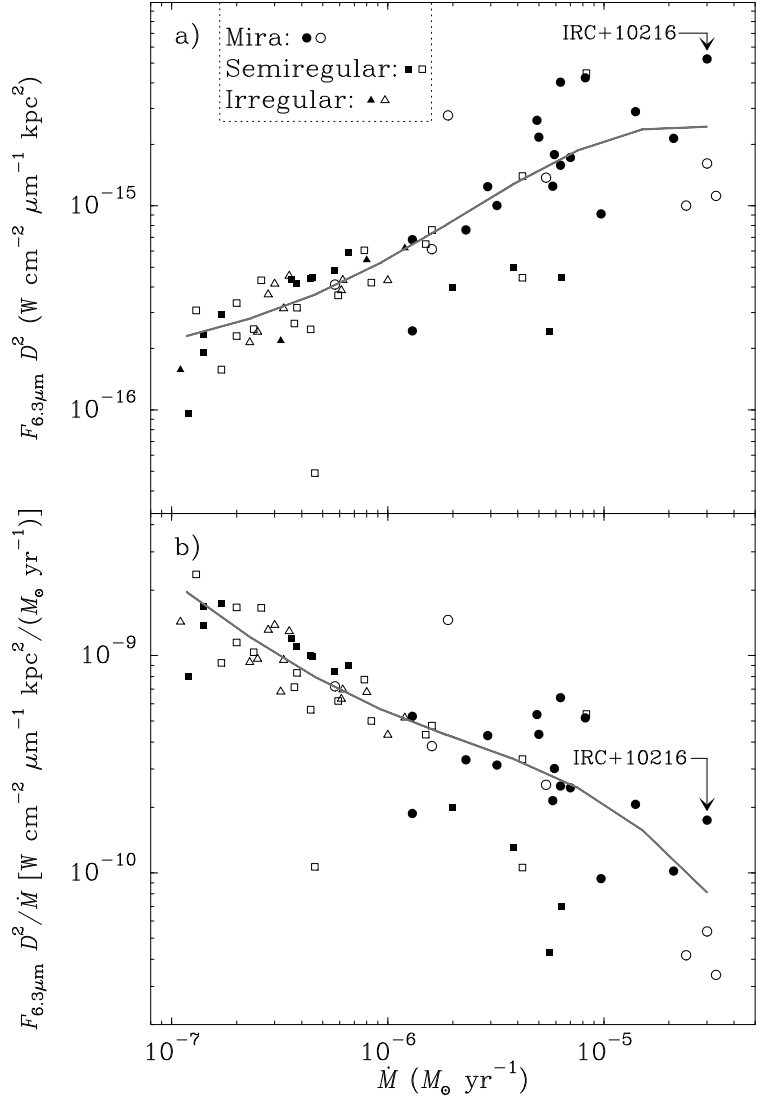}
\caption{a) Continuum flux densities at 6.3 $\mu$m, corrected for the 
distance to the sources, versus the mass loss rate for a sample of C-rich 
AGB stars. Filled symbols indicate sources for which the 6.3 $\mu$m
flux has been directly measured from ISO/SWS spectra
\citep{slo03}, and opened symbols indicate sources for which the 6.3 $\mu$m
flux has been estimated from the available 8.8 $\mu$m flux density (see text
for details). The grey line shows the fit to the general trend achieved
with our dust models. b) Same as in a) but divided by the mass loss rate.
The position of IRC+10216 is indicated in both diagrams.
\label{fig:cstarsflux} }
\end{figure}

The role of the mid-IR continuum emission in the excitation of \hdo\ 
in IRC+10216 should be also evaluated in other C-rich stars. To this end,
we show in Fig.~\ref{fig:cstarsflux}a the 6.3 $\mu$m mid-IR flux, corrected
for the distance, versus the mass loss rate for a sample of C-rich AGB 
stars. Most \Mdot\ and $D$ values have been taken from the compilation 
of \cite{gua06}. Distance estimates were mostly inferred from
HIPPARCOS astrometric measurements by \cite*{ber02}; the mass loss rates,
derived from millimetric molecular observations, were extracted from the
literature \citep[e.g.,][]{lou93} and subsequently corrected for the updated
distances by \cite{ber05}.
Filled symbols in  Fig.~\ref{fig:cstarsflux}a indicate sources for 
which the 6.3 $\mu$m flux has been directly measured from ISO/SWS spectra
\citep{slo03}\footnote{An Atlas of fully processed ISO/SWS spectra is 
publicly available at
http://isc.astro.cornell.edu/$\sim$sloan/library/swsatlas/atlas.html}. 
The ISO/SWS spectra has also allowed us to determine the 
6.3-to-8.8 $\mu$m flux density ratio as a function of \Mdot. 
This ratio, ranging from 1.25 for the highest \Mdot\ to 3.5 for the lowest 
\Mdot, has been used to estimate the 6.3 $\mu$m fluxes, shown with 
open symbols, for those sources not observed by ISO/SWS but 
for which the 8.8 $\mu$m flux is available \citep{gua06,mon98,lop93}. 
The 6.3 $\mu$m emission is responsible for the excitation of \hdo\ 
through the pumping of the $\nu_2=1$ state, and is thus a measure of the 
impact of the mid-IR emission on the excitation of \hdo. The plot shows 
that IRC+10216 has a high mid-IR intrinsic luminosity relative to other 
C-rich stars with similar \Mdot. In order to simulate the general trend
$L_{6.3\,{\rm \mu m}}-\Mdot$ found for the bulk of sources, we have generated 
dust models that differ in both \Mdot\ and the stellar luminosity ($L_*$). 
For the aim of simplicity, only $L_*$ has been varied as a function of \Mdot, 
while the rest of dust parameters remain equal to those adopted for IRC+10216 
(Table~\ref{tab:envel}). The grey line in Fig.~\ref{fig:cstarsflux}a shows
the fit found with $L_*=3.5\times10^3$ \Lsun\ for \Mdot$=1.2\times10^{-7}$
\Msun\ yr$^{-1}$, and $L_*=8\times10^3$ \Lsun\ for \Mdot$=3\times10^{-5}$
\Msun\ yr$^{-1}$.

Figure~\ref{fig:cstarsflux}b shows the 6.3 $\mu$m flux density, corrected
for the distance and divided by \Mdot, as a function of \Mdot. At any 
radial position of a given source, the mid-IR radiation density is expected 
to be proportional to $F_{6.3\,{\rm \mu m}}D^2$ and the H$_2$ density 
proportional to \Mdot, so that the quantity in the ordinates of 
Fig.~\ref{fig:cstarsflux}b is proportional to the radiative-to-collisional 
pumping rate ratio. If the kinetic temperature profile is not very
different from that adopted for IRC+10216, it is expected that \hdo\ in
those sources with $F_{6.3\,{\rm \mu m}}D^2/\dot{M}$ similar to or higher
than the value associated with IRC+10216 will be radiatively pumped. 
Figure~\ref{fig:cstarsflux}b shows that radiative pumping is expected to
dominate over collisional pumping at least for the bulk of sources with 
$\dot{M}\lesssim10^{-5}$ \Msun\ yr$^{-1}$.

We have used the above dust models to compute the expected \hdo\ emission
as a function of \Mdot. The key assumption for the \hdo\ models displayed
below is that the \hdo\ outflow rate (i.e. the total number of \hdo\ 
molecules in the envelope) is independent of \Mdot\ and equal to the value
found in IRC+10216 (which was required to explain the 557 GHz emission). 
That assumption is appropriate if the origin of the water vapor is the
vaporization of icy bodies, because the amount of comets orbiting a 
star is expected to be independent of \Mdot, whereas any other 
formation mechanism involves chemical reactions whose volume 
rates are proportional to the square
of the density. In the cometary hypothesis, the release of \hdo\ to the
outflow is still a function of the stellar luminosity, 
which is expected to increase with \Mdot, but also depends on other parameters
such as the evolutionary stage of the star after the onset of the TP-AGB phase
\citep{for01}, and thus a correlation between the \hdo\ outflow rate and 
\Mdot\ is uncertain but probably less pronounced than for any 
other formation mechanism. 

Another important model assumption is the parametrization
of the photodissociation radius for \hdo\ molecules, $R_{out}$, as a
function of the mass loss rate. Following \cite{net87}, we have assumed 
that $R_{out}$ scales as \Mdot$^{0.7}$, but normalized the values of
$R_{out}$ in such a way that $R_{out}=4\times10^{17}$ cm for 
\Mdot$=3\times10^{-5}$ \Msun\ yr$^{-1}$ (Willacy 2004, AC06). As a result of
this different normalization, our values for $R_{out}$ are one order of
magnitude larger than the radius of the OH shell calculated by \cite{net87}. 
For high \Mdot, our resulting values for $R_{out}$ are much
larger than the assumed $R_{int}=2.1\times10^{15}$ cm, 
and results are not sensitive to the adopted value for $R_{out}$. 
However, for \Mdot$\lesssim10^{-6}$ \Msun\ yr$^{-1}$, 
the finite size for the \hdo\ shell (e.g. $R_{out}/R_{int}\lesssim10$)
starts to lower the expected line fluxes, in particular that of 
the 557 GHz line. Since \cite{net87} obtain values for $R_{out}$ 
that are still much lower than ours, one should consider with 
caution the \hdo\ fluxes predicted for \Mdot$\lesssim10^{-6}$ 
\Msun\ yr$^{-1}$. Nevertheless, we note that the photodissociation rate 
in the unshielded interstellar radiation field ($5.9\times10^{-10}$
s$^{-1}$) yields, for a 15 km s$^{-1}$ flow, a lengthscale  of
$\sim2\times10^{15}$ cm.  For $R_{int}=2\times10^{15}$ cm, this implies a
minimum value for $R_{out}$ of $\sim4\times10^{15}$ cm for any \Mdot\ 
however small, unless the radiation field is stronger than the 
average value. For the lowest \Mdot$=1.2\times10^{-7}$ \Msun\ yr$^{-1}$
considered here, the adopted $R_{out}$ is $8.4\times10^{15}$ cm.

\begin{figure}
\epsscale{0.6}
\plotone{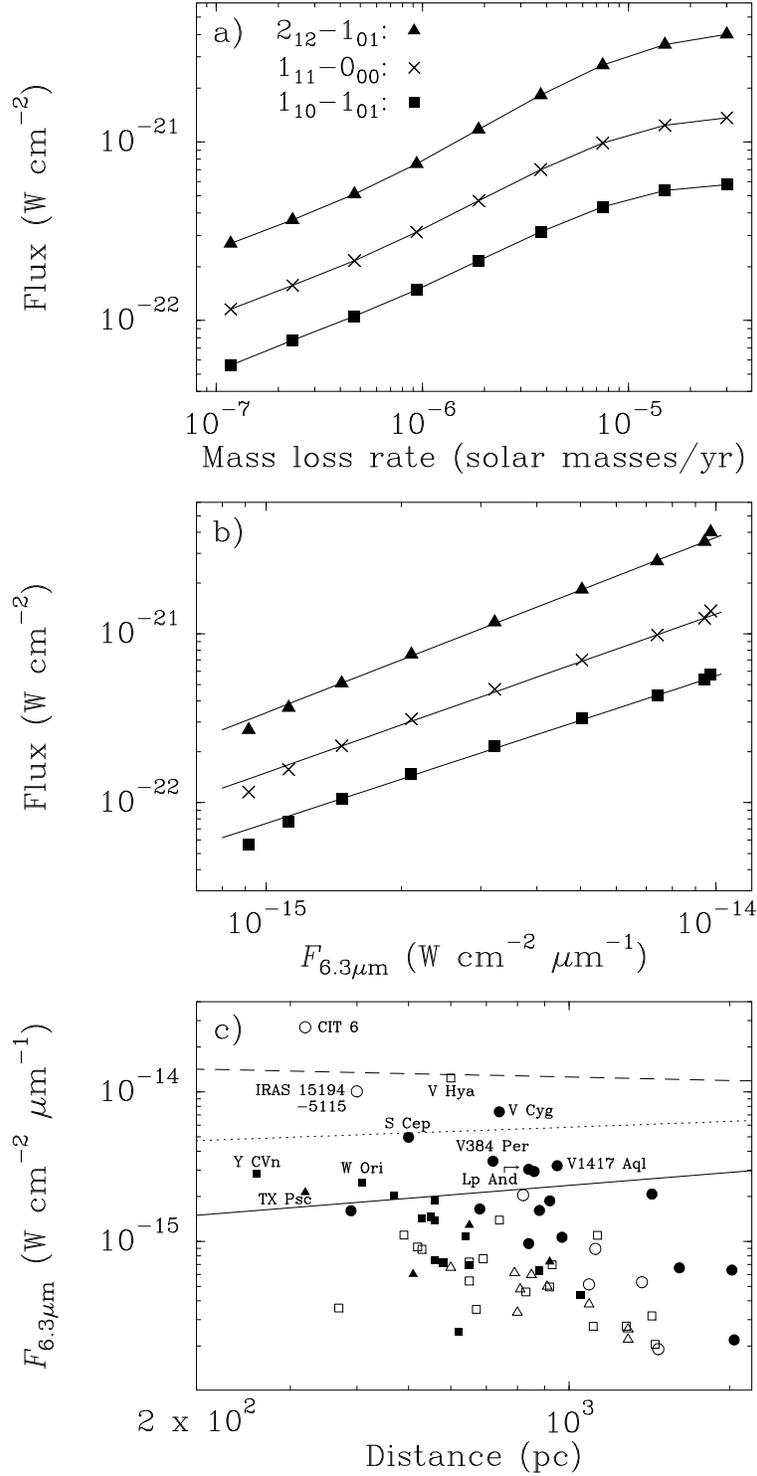}
\caption{Calculated \hdo\ line fluxes vs. \Mdot\ (a),
and vs. the 6.3 $\mu$m continuum flux (b), where the lines are fits of
the form $F_{{\rm line}}\propto F_{{\rm 6.3\mu m}}^b$; 
results are given for a source at $D=0.5$ kpc. 
c) Observed continuum fluxes at 6.3 $\mu$m 
of C-rich AGB stars in function of the distance 
(see Fig.~\ref{fig:cstarsflux} for the meaning of symbols). The
lines show the 6.3 $\mu$m continuum flux required to detect
the ground-state lines in 1.5 hours of observing time with HSO-HIFI,
if the \hdo\ outflow rate is the same as that in IRC+10216 (solid: \t110101;
dotted: \t111000; dashed: \t212101). Some of the stars that would
be detectable in such a case in the \t110101\ o-\hdo\ line at 557 GHz 
are labelled.
\label{fig:flujosmdotf6um2}}
\end{figure}

Figure \ref{fig:flujosmdotf6um2} shows the expected fluxes of the three
ground-state lines of \hdo\ as a function of a) \Mdot, and b) the 6.3 $\mu$m 
continuum flux; results are given for a source at $D=0.5$ kpc. We find
that the excitation is dominated by absorption of mid-IR photons; 
only for the highest \Mdot\ does collisional 
excitation account for at most $20\%$ of the predicted flux in 
the ground-state lines. Therefore, a tight correlation between line 
fluxes and the 6.3 $\mu$m continuum flux density is found in 
Fig.~\ref{fig:flujosmdotf6um2}b, where fits of the form 
$F\propto F_{6.3\mu m}^b$ are displayed.
The slope $b$ is $\approx1$ for the 1670 GHz
line, but somewhat lower for the 557 and 1113 GHz lines (0.87 and 0.94,
respectively) due to slight opacity effects in the vibrational lines 
responsible for the radiative pumping. 
The slopes would approach to unity for a microturbulent velocity higher 
than the assumed value of $0.65$ \kms. For low \Mdot, the data points are
slightly below the fitted lines as a result of the \hdo\ photodissociation 
effects mentioned above; stronger departures from the fitted lines are 
expected for lower values of $R_{out}$.

According to the current estimates
of the HSO-HIFI sensitivities (Table~\ref{tab:h2otrans}), the 557 GHz line
will be detected in 1.5 hours at more than 5$\sigma$ level if the line 
flux is higher than $1.3\times10^{-22}$ W cm$^{-2}$ (after coadding the two
polarization modes). From Fig.~\ref{fig:flujosmdotf6um2}b, this sensitivity 
limit translates into a 6.3 $\mu$m flux density of $1.7\times10^{-15}$ 
W cm$^{-2}$ $\mu$m$^{-1}$ at $D=0.5$ kpc, which can be in turn adapted to 
any other distance provided that both the line and continuum 
fluxes vary as $D^{-2}$. Figure~\ref{fig:flujosmdotf6um2}c shows the
6.3 $\mu$m fluxes for the sample of C-rich AGB stars, with the solid line
indicating the 6.3 $\mu$m flux density required to detect the 557 GHz line
according to the quoted sensitivity limit. {\em All sources above the solid 
  line would be detected with 1.5 hours of observing time in the 557 GHz line
  if the \hdo\ outflow rate were equal or higher than the value
  found in IRC+10216, and if the photodissociation radius of the \hdo\
  molecules remains sufficiently large relative to $R_{int}$}. 
The sensitivity limits for the 1113 and 1670 GHz lines are also 
displayed, and show that the detection of these lines require 
significantly higher 6.3 $\mu$m continuum fluxes. We conclude that the 
557 GHz line provides the most sensitive means of searching
for \hdo\ in C-rich AGB stars other than IRC+10216 with HSO.

\section{Discussion} \label{sec:discussion}

\subsection{Model uncertainties}

Our models for IRC+10216 predict that, whatever the region where \hdo\ is
formed or released to the outflow, \hdo\ is radiatively excited, which
implies that model results are independent of the gas temperature and density
profiles. This result relies on the extrapolation to higher $T_k$ of the
collisional rates given by \cite{phi96}, which will require further
confirmation from new estimates of \hdo-H$_2$ collisional rates at higher
temperatures. Nevertheless, we find that collisional rates 
would have to be about
one order of magnitude higher than estimated to compete efficiently with 
the radiative rates in the innermost regions of the envelope, which seems 
somewhat implausible.

A relatively uncertain parameter in our models is the assumed 
distance to IRC+10216, $D=170$ pc. It has been proposed that the distance 
may be substantially smaller, $D=100-150$ pc \citep*{zuc86,gro92}.
At 170 pc, the inferred stellar luminosity is $L_*\approx2\times10^4$ \Lsun,
which results in $L_*\approx10^4$ \Lsun\ at $D=120$ pc, a value more similar
to the inferred luminosities of other C-rich AGB stars with similar \Mdot\
\citep*{sch06}. If $D=120$ pc, both \Mdot\ and the radiation density at 6
$\mu$m are a factor of 2 lower than assumed, so that the 
radiative-to-collisional pumping rate ratio still remains unchanged. 
Since the closer proximity and weaker 6 $\mu$m radiation density
have opposite effects on the \hdo\ outflow rate required to match the SWAS 
557 GHz flux, the latter will decrease by less than a factor 2, and
thus the expected \hdo\ abundance will be a factor of $<2$ higher than 
in Table~\ref{tab:h2omodels}. Therefore, we conservatively estimate an 
o-\hdo\ abundance in the range $(0.5-1)\times10^{-7}$ for 
$R_{int}=2\times10^{15}$ cm. Concerning the line ratios to be 
observed by HSO, we expect values similar to those obtained at $D=170$ 
pc if the line emission remains unresolved, i.e. for low values of 
$R_{int}$. For high values of $R_{int}$, beam effects will be more 
important at $D=120$ pc, diminishing the low-excitation line fluxes 
relative to the 557 GHz line.

A potentially more important source of uncertainty is the assumption of 
spherical symmetry in our models. Both the molecular and dust emission 
from IRC+10216 show evidence for departures from a smooth distribution 
in a spherically symmetric outflow; incomplete, discrete shells or arcs
and clumpy structures are instead observed on a wide range of distances 
to the star \citep*[e.g.][]{luc99,mau99,fon03,lea06}. Even more important, 
the OH line shapes observed in IRC+10216 strongly suggest an asymmetric 
distribution of OH in the outer regions of the envelope \citep{for03}, 
thus also suggesting an asymmetric distribution of the parent \hdo\ molecule. 
These asymmetries may alter to some extent the \hdo\ line flux ratios 
calculated with the use of our spherically symmetric approach. While in 
spherical symmetry the emission from any line is isotropic, in filamentary 
structures or slabs of velocity-coherent gas the optically thick lines 
radiate preferentially in the direction perpendicular to the two faces of 
the sheet \citep*{eli89}, whereas the emission from thinner lines 
will approach a more isotropic behavior. Future HSO observations will 
show the importance of the observed morphological complexity on the line 
fluxes by showing whether the different \hdo\ line flux ratios are 
consistent with a single value of $R_{int}$, or indicate a range of values.

Finally, the models assume a water shell with uniform \hdo\ abundance and 
sharp inner and outer radii; however, a finite abundance gradient obviously
takes place at both the \hdo\ formation and dissociation regions, and
variations of the \hdo\ abundance across the shell are also possible. 
These effects may also alter to some extent the expected line flux 
ratios. Uncertainties in the outer radius of the \hdo\ shell may
also affect the expected fluxes from stars with low mass loss rates.

\subsection{Water formation at inner or intermediate radii}

Both the cometary and Fischer-Tropsch catalysis hypothesis predict \hdo\ 
to be released or formed at intermediate radii of {\it a few} 
$\times10^{15}$ cm, and although no specific mechanism has been proposed 
for \hdo\ formation at the innermost regions ($R_{int}<10^{15}$ cm), this 
possibility cannot be rejected. The models shown in 
section~\ref{sec:predictions} will permit us to discriminate, from HSO 
observations of IRC+10216, if there are significant amounts of \hdo\ in 
the innermost regions through the observation of mid-excitation \hdo\ 
transitions. More difficult will be a priori to discriminate between the 
cometary and Fischer-Tropsch catalysis propositions. Both predict similar 
values for $R_{int}$; nevertheless, if $R_{int}$ were found to be 
significantly higher than $2\times10^{15}$ cm, the release of \hdo\ from 
comets could be favoured, unless some additional mechanism were found to 
shift $R_{int}$, within the Fischer-Tropsch catalysis framework 
\citep{wil04}, outwards in the envelope. The search for water emission at 
557 GHz in C-rich AGB stars other than IRC+10216 will also favour one of 
the two hypotheses: since IRC+10216 is at the high end of mass loss rates 
from C-rich stars, and the efficiency of Fischer-Tropsch catalysis to 
form \hdo\ molecules is expected to decrease with diminishing \Mdot, one 
would expect in such a case a rate of detection significantly lower than 
that quantified in section~\ref{sec:crich} for the cometary hypothesis,
and one would expect a pronounced decline of the \hdo\ abundance with 
diminishing \Mdot.

The ISO/LWS spectrum of IRC+10216 \citep{cer96} shows an emission feature
at 179.5 $\mu$m, coincident with the wavelength of the \t212101\ o-\hdo\
transition, with a flux of $\approx8\times10^{-20}$ W cm$^{-2}$. This flux is
very similar to that computed for the quoted line in model $B$ 
($R_{int}=2.1\times10^{15}$ cm). However, the far-infrared spectrum of
IRC+10216 shows vibrationally excited rotational emission of HCN, and the 
combined emission of the 
$\nu_1=1$ $J=19\rightarrow18$ and $\nu_3=1$ $J=19\rightarrow18$
HCN lines, both emitting at 179.5 $\mu$m, is expected to be also comparable
to the measured line flux at 179.5 $\mu$m \citep{cer96}.  Given
the uncertainties inherent to the HCN model in \cite{cer96}, where the
excited vibrational states are assumed to be thermalized, and given
the high density
of spectral lines in the IRC+10216 far-infrared spectrum, which makes it
difficult to establish the contribution from the $\nu_1=1$ and $\nu_3=1$
rotational lines to the spectrum, the relative 
contribution from HCN and \hdo\ to the observed spectral feature 
is quite uncertain. On the other hand, the expected flux of the 
\t303212\ line at 174.6 $\mu$m in model $B$ is $\approx4\times10^{-20}$ 
W cm$^{-2}$, below the 3-$\sigma$ upper limit derived for that line from the 
ISO/LWS spectrum. These considerations suggest weakly that 
\hdo\ in IRC+10216 is formed or released at radial distances 
$R_{int}\gtrsim2\times10^{15}$ cm.

Our models show that the required \hdo\ abundance for $R_{int}$ in the range
$(2-5)\times10^{15}$ cm is $(0.5-1)\times10^{-7}$ relative to H$_2$, which 
makes the requirements previously demanded for the cometary 
and Fischer-Tropsch catalysis hypothesis to work in IRC+10216 less restrictive 
\citep{for01,wil04}. The water outflow rate is $(0.5-1)\times10^{-5}$ 
M$_{\oplus}$ yr$^{-1}$, which yields, within the framework of vaporization 
of icy bodies, a required total initial ice mass of $(0.5-10)$ M$_{\oplus}$ 
for \Mdot(\hdo)/$M_0({\rm ice})$ in the range $10^{-5}-10^{-6}$ yr$^{-1}$
\citep{for01}. On the other hand, the Fischer-Tropsch catalysis on metallic
grains will require a density of iron grains relative to total H
nuclei of $n_g({\rm Fe})/n_{{\rm H}}=(0.5-1)\times10^{-13}$ to 
explain the observed \hdo\ 557 GHz emission \citep{wil04}.

\subsection{Water formation in the outermost layers}
\label{sec:radassoc}

Multitransition HSO observations of \hdo\ in IRC+10216 will easily establish
whether or not water is formed in the external layers of the envelope; 
the predicted
fluxes in Fig.~\ref{fig:flujosrint2} and the HSO-HIFI sensitivities in 
Table~\ref{tab:h2otrans} indicate that only the \t110101\ o-\hdo\ line,
and possibly the \t111000\ and \t202111\ p-\hdo\ lines, are detectable in one
hour of observing time for $R_{int}\gtrsim4\times10^{16}$ cm.

For high $R_{int}$, the expected flux in the 557 GHz line can 
be obtained analytically. Since essentially all o-\hdo\ molecules are in 
the ground $1_{01}$ level (Fig.~\ref{fig:pump557ghz}a), the o-\hdo\ \t110101\ 
line flux is derived from
the radiative pumping rate given in eq.~(\ref{gammar}) after multiplying 
$\Gamma_r$ by 1.35 to account for the two radiative pumping 
routes that are relevant at high distances from the star 
(Fig.~\ref{fig:pump557ghz}b):
\begin{eqnarray}
F(1_{10}\rightarrow1_{01})  =  1.4\times10^{-21} \times
\left( \frac{4\times10^{17} \, {\rm cm}}{R_{out}} \right) \nonumber \\ 
\times 
\left( \frac{R_{out}}{R_{int}} -1 \right)  \times 
\left( \frac{X({\rm o-H_2O})}{5\times10^{-7}} \right) \,\, 
{\rm \frac{W}{cm^{2}}},
\label{fluxanal}
\end{eqnarray}
where $X$ is the abundance relative to H$_2$. Equation~(\ref{fluxanal}) is 
independent of the assumed distance $D$ to the star because, in order to 
match the observed mid-IR continuum, $\Gamma_r\propto D^{2}$ (eq.~\ref{eqj}); 
it assumes a water shell with sharp edges at $R_{int}$ and $R_{out}$ and 
constant \hdo\ abundance; it ignores slight opacity effects in the 
ro-vibrational lines (section~\ref{sec:110101}) as well as beam effects 
(both slightly raise the required abundance), and overestimates by less
than 20\% the 557 GHz line flux obtained in model $C$. In the 
limit $R_{out}\rightarrow\infty$, and using 
$F(1_{10}\rightarrow1_{01})=10^{-20}$ W cm$^{-2}$ \citep{mel01}, 
eq.~(\ref{fluxanal}) gives
\begin{equation}
\chi({\rm H_2O}) \gtrapprox 2.4\times10^{-7} 
\times \left( \frac{R_{int}}{4\times10^{16} \, {\rm cm}} \right),
\label{fluxanallim}
\end{equation}
where $\chi({\rm H_2O})$ is here the \hdo\ (ortho+para) abundance relative
to H nuclei. Equation~\ref{fluxanallim} gives the sharp-inner edge, lower limit
for the \hdo\ abundance required to account for the observed 557 GHz line flux,
as a function of $R_{int}$. 
AC06 have reported an abundance of $\chi({\rm H_2O})\sim10^{-7}$ to
account for the observed 557 GHz line flux in IRC+10216. However,
the quoted abundance would imply an inner radius of
$R_{int}\lessapprox2\times10^{16}$ cm, but the \hdo\ abundance shown by 
AC06 (their Fig. 7) decreases sharply at $r<4\times10^{16}$ cm.
Based on detailed modelling, we indicate that the $\chi({\rm H_2O})$ 
profile given by AC06 has to be shifted up by a factor of $\approx2.5$ 
to account more accurately for the 557 GHz line flux measured by SWAS.

In the model proposed by AC06, atomic oxygen is produced in a shell by 
the photodissociation of CO (and particularly the $^{13}$CO isotopologue, 
which shields itself far less effectively than $^{12}$CO.)  Since the 
temperature is low ($\sim 10$~K) within the shell where the atomic oxygen 
abundance is significant, the neutral-neutral reaction sequence 
$$\rm O + H_2 \rightarrow OH + H$$
$$\rm  OH + H_2 \rightarrow H_2O + H$$
is very slow --and therefore is negligible as a source of H$_2$O-- the first 
reaction being endothermic and the second --although exothermic-- 
possessing a substantial activation energy barrier.

AC06 therefore proposed the radiative association reaction
$$\rm O + H_2 \rightarrow H_2O + \gamma$$ 
as an alternative source of H$_2$O.  To match the SWAS- and ODIN-observed 
water line fluxes, AC06 had to posit that this reaction is relatively 
rapid at low temperature, with a rate coefficient 
$k_{ra} \rm \sim 10^{-15}\,cm^3\,s^{-1}$. 
According to our estimation above, $k_{ra}$ has to be a
factor $\approx2.5$ higher than this value to account for the 
measured \t110101\ \hdo\ flux. Nevertheless, and even with the use of the
$k_{ra}$ estimation given by AC06, we find that this large reaction
rate coefficient for the radiative association of O and
H$_2$ is inconsistent with observations of H$_2$O and OH in 
at least one translucent molecular cloud, for which sufficient data
exist to disipate any significant ambiguity: observations of the cloud
along the sight-line to HD 154368 --carried out by \cite{spa98} 
with the use of the Goddard High Resolution Spectrograph (GHRS) on the 
{\it Hubble Space Telescope} (HST) -- yield a 3~$\sigma$ upper limit on 
the water abundance that lies almost two orders of magnitude below the 
value that would obtain were $k_{ra}$ as large as 
$\rm 10^{-15}\,cm^3\,s^{\rm -1}$. The factor of discrepancy would rise
above $150$ for $k_{ra} \rm \approx 2.5 \times 10^{-15}\,cm^3\,s^{-1}$.

The best-fit model for the HD 154368 cloud obtained by 
\cite{spa98} posits a plane parallel cloud of total 
visual extinction $A_V = 2.65$~mag in which the density of H nuclei 
is $n_{\rm H} = 325\, \rm cm^{-3}$ \citep[this is probably a lower
limit at the cloud center, as the density inferred from the CO
$J=1\rightarrow0/3\rightarrow2$
ratio is $\sim10^3$ cm$^{-3}$; see][]{van91}
and the external ultraviolet 
radiation field is 3 times mean interstellar value given by
\cite{dra78}: $I_{UV} = 3$.  Under these conditions, the destruction 
of H$_2$O is dominated by photodissociation at a rate 
$\zeta = 5.9 \times 10^{-10}{\rm s}^{-1} \times  I_{UV} \times 
(\exp\{-1.7 A_{V1}\} + \exp\{-1.7 A_{V2}\})/2$ 
\citep*{let00}\footnote{We note that the factor of 1.7 in the
exponential, widely used in the literature, yields \hdo\ photodissociation
rates at the midplane of the plane-parallel cloud that are higher than
the values reported by \cite{rob91} by factors of 2.2 and 10 for
$A_V^{tot}=1$ mag and $A_V^{tot}=10$ mag, respectively; therefore,
our estimation for the $N({\rm H_2O})/N({\rm O})$ ratio predicted by 
the radiative association of O and H$_2$ is probably a conservative 
lower limit.}, 
where $A_{V1}$ is the visual 
extinction to one cloud surface and $A_{V2} = 2.65 - A_{V1}$ is the 
extinction to the other.  At the cloud center, the water 
photodissociation rate is $1.9 \times 10^{-10}\,\,{\rm s}^{-1}$.

\begin{figure}
\epsscale{0.9}
\plotone{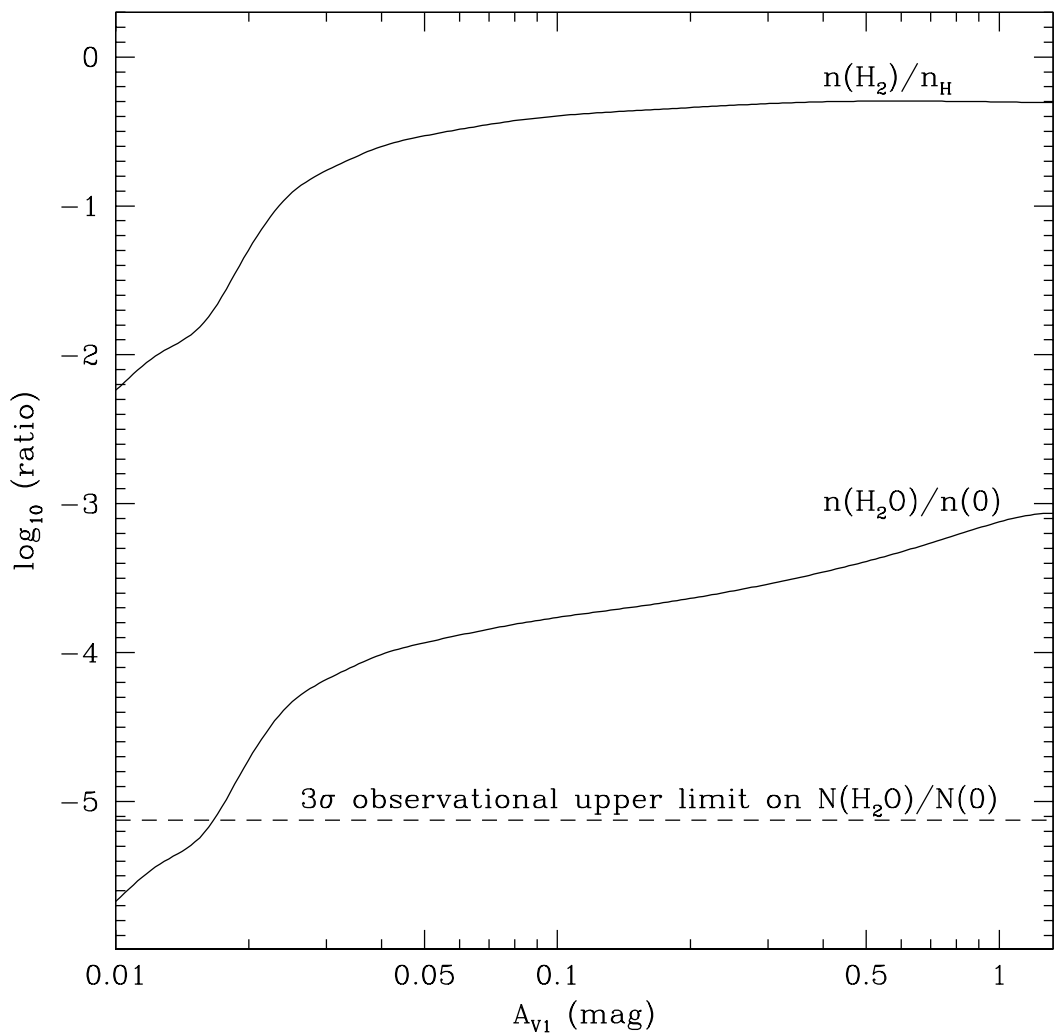}
\caption{
Predicted $n({\rm H_2}) / n_{\rm H}$ ratio for the best-fit 
\cite{spa98} model, together with the 
$n({\rm H_2O})/n({\rm O})$ ratio that would result if $k_{ra}$ were 
$\rm 10^{-15}\,cm^3\,s^{-1}$.  The dashed line shows the 3~$\sigma$ 
upper limit on the observed $N({\rm H_2O})/N({\rm O})$ ratio.
\label{fig:nh2ono}}
\end{figure}

For a radiative association rate of $k_{ra}$, 
the ratio of water vapor to atomic oxygen is therefore given by the 
expression
$$n({\rm H_2O})/n({\rm O}) = k_{ra} n_{\rm H} f_{\rm H_2}/ \zeta,$$
where $f_{\rm H_2} \equiv n({\rm H_2}) / n_{\rm H}$.  At the cloud center, 
the cloud is almost fully molecular, with $f_{\rm H_2} \sim 0.5$, and the 
above equation yields 
$n({\rm H_2O})/n({\rm O}) = 8.7 \times 10^{-4} 
\,(k_{ra}/\rm 10^{-15}\,cm^3\,s^{\rm -1})$.  In Fig.~\ref{fig:nh2ono}, we 
show the predicted $n({\rm H_2}) / n_{\rm H}$ ratio for the best-fit 
\cite{spa98} model, together with the 
$n({\rm H_2O})/n({\rm O})$ ratio that would result if $k_{ra}$ were 
$\rm 10^{-15}\,cm^3\,s^{-1}$.  
The model predicts hydrogen to be predominantly in molecular form, in
agreement with results by \cite{sno96}.
Averaging the $n({\rm H_2O})/n({\rm O})$ ratio over the entire 
sight-line, we obtain a column density ratio 
$N({\rm H_2O})/N({\rm O}) = 5.3 \times 10^{-4} 
\,(k_{ra}/\rm 10^{-15}\,cm^3\,s^{\rm -1})$.

The atomic oxygen column density along the HD 154368 sight-line is 
$N({\rm O}) = 1.2 \times 10^{18}\,\,{\rm cm^{-2}}$
(Snow et al. 1996, from absorption line observations of the 
1355 \AA\ OI] line).
Based on a search for the $C^1B_1 - X^1A_1$ band of water vapor near 
1240 \AA, \cite{spa98} obtained 3~$\sigma$ upper limit on the 
water column density of $9 \times 10^{12}\,\,{\rm cm^{-2}}$, corresponding to 
$N({\rm H_2O})/N({\rm O}) = 7.5 \times 10^{-6}$.  This 3~$\sigma$ upper 
limit lies a factor 70 below the value that we would obtain were 
$k_{ra}$ equal to $10^{-15}\rm \,cm^3\,s^{-1}$ as AC06 suggested, and places a 
3~$\sigma$ upper limit of $1.4 \times 10^{-17}\rm  \,cm^3\,s^{-1}$ on 
$k_{ra}$.

AC06 suggested that the freeze out of oxygen onto grain mantles could 
diminish the water vapor abundance in molecular clouds, as proposed by 
\cite{ber00} to explain the low H$_2$O abundances measured by 
SWAS in dense clouds.  Ice absorption line observations of diffuse/translucent 
sight-lines, however, indicate that water ice is generally present only in 
clouds of $A_V \ge 3$ \citep{whi01}. Furthermore,
in the specific case of HD 154368 under present consideration, 
the atomic oxygen is known from direct measurement to be 
$1.2 \times 10^{18}\,\,{\rm cm^{-2}}$.  The 
column density of H nuclei along this sight-line, 
$N_{\rm H} = N({\rm H}) + 2 N({\rm H}_2)$, has been measured to be 
$4.2 \times 10^{21}\,\,\rm cm^{-2}$ \citep{sno96}, so the mean 
line-of-sight oxygen abundance is 
$n({\rm O})/n_{{\rm H}} \sim 3 \times 10^{-4}$, 
a value that is entirely consistent with the abundances measured along 
diffuse sight-lines \citep{mey98} and inconsistent with a significant 
depletion of oxygen onto ice mantles.

In summary, the upper limit on the water vapor abundance observed towards 
HD 154368 definitively rules out a rate coefficient for the radiative 
association reaction $\rm O + H_2 \rightarrow H_2O + \gamma$ that is large 
enough to explain --in the context of the AC06 model-- the H$_2$O 
\t110101\ line strength measured by SWAS toward IRC+10216. 
If HSO observations would indicate that \hdo\ is formed in the external
layers of IRC+10216, an explanation other than the radiative association
proposed by AC06 would be required to avoid incompatibilities with
observations toward HD 154368.

\section{Summary} 

Our radiative transfer models for \hdo\ in IRC+10216 and other C-rich AGB
stars reveal: \\
1) \hdo\ in the envelope of IRC+10216 is primarily excited through absorption 
of photons in the $\nu_2$ band at 6 $\mu$m and subsequent decay to the ground
vibrational state. \\ 
2) The o-\hdo\ abundance relative to H$_2$ required to account for the 557 
GHz \t110101\ o-\hdo\ line observed in IRC+10216 is in the range 
$(0.5-5)\times10^{-7}$, depending on the inner radius of the \hdo\ shell. \\
3) Multitransition \hdo\ observations with the Herschel Space Observatory
will allow us to establish the spatial distribution of \hdo\ molecules in
IRC+10216, in particular the inner radius of the \hdo\ shell, thus
discriminating among the different hypothesis proposed for the origin
(formation or release) of \hdo. \\
4) A number of other C-rich AGB stars with relatively strong 6 $\mu$m
continuum flux are expected to be detectable in the \t110101\ o-\hdo\ line 
with HSO, if the \hdo\ outflow rate in those sources is similar to that 
found in IRC+10216. \\
5) The relatively low \hdo\ abundance required to explain the \t110101\
o-\hdo\ emission in IRC+10216 makes less restrictive the requirements
previously reported for the vaporization of icy bodies or 
Fischer-Tropsch catalysis on the surfaces of metallic grains to work.
If \hdo\ were formed in the external layers of IRC+10216, we find that
an explanation for the \hdo\ formation other than the previously 
reported radiative association $\rm O + H_2 \rightarrow H_2O + \gamma$ 
is required to avoid conflict with observations of the translucent cloud
towards HD 154368.

\acknowledgments
D.A.N. gratefully acknowledges the support of grant NAG 5 - 13114 from
NASA's Long Term Space Astrophysics research program.
This research has made use of NASA's Astrophysics Data System.

\clearpage


\begin{deluxetable}{cc}
\tablecaption{Model parameters for IRC+10216  \tablenotemark{a} 
\label{tab:envel}}
\tablewidth{0pt}
\tablehead{
\colhead{Parameter} & \colhead{Value} } 
\startdata
Gas velocity field & $v(r)=14.5\sqrt{1-0.95\frac{R_i}{r}}$ km s$^{-1}$ \\
Turbulent velocity & $v_t=0.65$ km s$^{-1}$ \\
H$_2$ density & $n({\rm H_2})=\frac{3.11\times10^{7}}{r_{15}^{2}} 
\frac{14.5\,\,{\rm km s^{-1}}}{v(r)} \,\, {\rm cm^{-3}}$ \\
Gas temperature & $T_k(r)=12\left(\frac{90}{r_{15}}\right)^{0.72}$ K if
$r_{15}<110$ \\
& $T_k(r)=10$ K if $r_{15}>110$ \\
Dust inner radius & $R_i=4.5\times10^{14}$ cm \\
Dust outer radius & $R_o=4.0\times10^{17}$ cm \\
Stellar radius & $R_*=8.0\times10^{13}$ cm \\
Stellar temperature & $T_*=2050$ K \\
Spectral index & $\beta=1.3$ if $\lambda>20$ $\mu$m \\
& $\beta=1$ if $\lambda<20$ $\mu$m \enddata
\tablenotetext{a}{$r$ is the distance to the star in cm, and 
$r_{15}=\frac{r}{10^{15} {\rm cm}}$.}
\end{deluxetable}
 


\begin{deluxetable}{cccc}
\tablecaption{Inner and outer radii for the \hdo\ shell in different
models, and derived o-\hdo\ abundances 
\label{tab:h2omodels}}
\tablewidth{0pt}
\tablehead{
\colhead{Model} & \colhead{$R_{int}$ (cm)} & \colhead{$R_{out}$ (cm)} 
& \colhead{$X$(o-\hdo)\tablenotemark{a}} 
}
\startdata
$A$ & $4.5\times10^{14}$ & $4.0\times10^{17}$ & $4.1\times10^{-8}$ \\
$B$ & $2.1\times10^{15}$ & $4.0\times10^{17}$ & $5.4\times10^{-8}$ \\
$C$ & $4.3\times10^{16}$ & $4.0\times10^{17}$ & $5.1\times10^{-7}$ 
\enddata
\tablenotetext{a}{o-\hdo\ abundances are given relative to H$_2$, and
an ortho-to-para \hdo\ abundance ratio of 3:1 is assumed in all models.}
\end{deluxetable}


\begin{deluxetable}{cccc}
\tablecaption{H$_2$O transitions, frequencies, and adopted
HSO-HIFI half-power beamwidths and sensitivities
\label{tab:h2otrans}}
\tablewidth{0pt}
\tablehead{
\colhead{Transition} & \colhead{Frequency} & 
\colhead{HSO-HIFI beam\tablenotemark{a}} 
& \colhead{Sensitivity\tablenotemark{b}} \\
& (GHz) & ($''$) & ($10^{-22}$ W cm$^{-2}$)
}
\startdata
\t110101 & 556.9  & 41.0 & 2.2 \\
\t312303 & 1097.4 & 21.0 & 12 \\
\t312221 & 1153.1 & 18.5 & 29 \\
\t321312 & 1162.9 & 18.5 & 30 \\
\t523514 & 1410.6 & 14.5 & 45 \\
\t221212 & 1661.0 & 14.5 & 61 \\
\t212101 & 1669.9 & 14.5 & 62 \\
\t303212 & 1716.8 & 12.5 & 56 \\
\t532523 & 1867.7 & 12.5 & 65  \\
\t211202 & 752.0  & 29.0 & 4.1 \\
\t202111 & 987.9  & 21.0 & 9.2 \\
\t111000 & 1113.3 & 21.0 & 12 \\
\t422413 & 1207.6 & 18.5 & 32 \\
\t220211 & 1228.8 & 18.5 & 35 \\
\t413404 & 1602.2 & 14.5 & 55 \\
\t633624 & 1762.0 & 12.5 & 60 \\
\t624615 & 1794.8 & 12.5 & 64 
\enddata
\tablenotetext{a}{Adopted from 
http://www.ipac.caltech.edu/Herschel/hifi/hifi.shtml}
\tablenotetext{b}{Calculated using HSPOT version 1.9.1; sensitivities
correspond to 5$\sigma$ in 1 hr with a spectral resolving 
power of $R=10^{4}$}
\end{deluxetable}

\clearpage


\end{document}